\documentclass[12pt]{article}

\usepackage{graphicx}
\usepackage{caption}

\usepackage{amsmath}
\usepackage[dvipsnames]{xcolor}

\usepackage{lineno}

\def\beq{\begin{equation}}
\def\eeq{\end{equation}}
\def\bea{\begin{eqnarray}}
\def\eea{\end{eqnarray}}

\title{Quantum machine learning using atom-in-molecule-based fragments selected on-the-fly}

\author
{Bing Huang and O. Anatole von Lilienfeld$^{\ast}$\\
\\
\normalsize{Institute of Physical Chemistry and National Center }\\
\normalsize{for Computational Design and Discovery of Novel Materials (MARVEL),}\\
\normalsize{Department of Chemistry, University of Basel, }\\
\normalsize{Klingelbergstrasse 80, 4056 Basel, Switzerland}\\
\\
\normalsize{$^\ast$To whom correspondence should be addressed; E-mail:  anatole.vonlilienfeld@unibas.ch.}
}

\date{}

\begin{document} 





\maketitle


\begin{abstract}
First principles based exploration of chemical space deepens our understanding of chemistry, and might help with the design of new molecules, materials or experiments. 
Due to the computational cost of quantum chemistry methods and the immens number of theoretically possible 
stable compounds comprehensive in-silico screening remains prohibitive. 
To overcome this challenge, we combine atoms-in-molecules based fragments, dubbed ``amons'' (A), 
with active learning in transferable quantum machine learning (ML) models.
The efficiency, accuracy, scalability, and transferability of resulting AML models is 
demonstrated for important molecular quantum properties, such as energies, forces, atomic charges NMR shifts, polarizabilities, 
and for systems including organic molecules, 2D materials, water clusters, Watson-Crick DNA base-pairs and even ubiquitin. Conceptually, the AML approach extends Mendeleev's table to effectively account for chemical environments, which allows the systematic reconstruction of many chemistries from local building blocks. 
\end{abstract}

The basic concept of fundamental building blocks, determining the behaviour of matter through their specific combinations, 
has had a profound impact on our understanding of particle physics (quarks)~\cite{feynman1963}, 
electronic structure of atoms and molecules (elemental particles)~\cite{martin2004}, 
or proteins and genetic codes (amino and nuclear acids)~\cite{biology2011}. 
Within the molecular sciences, the scalability and transferability among functional groups has
enabled synthetic chemists to reach remarkably precise control over intricate complex atomistic processes.

Recently developed ML models of quantum properties throughout chemical compound space, 
by contrast, suffer from severe limitations because their domain of applicability has to resemble the data used for 
training~\cite{CM,ML4Polymers_Rampi2013,ML4Crystals_Wolverton2014,Alan_OLED2015,Felix2016,Sandip2016,DTNN2017,GrossmannCNN2018,ANI,SchNet} \textcolor{black}{and thus the transferability features of functional groups in molecules cannot be fully reaped and utilized for accurate predictions of larger molecular systems}.
\textcolor{black}{More specifically, conventional ML-protocols rely on some pre-defined dataset, 
whose origin typically suffers from severe bias and which is sub-sampled at random (or through active 
learning~\cite{Shapeev_ActiveLearning, Ceriotti_AutoKernelSelect}) 
for training and testing. This procedure entails at least two inherently challenging drawbacks: 
(i) Scalability: When the number of compositional elements and/or size of query systems increases, 
the composition and size of the training compounds must follow suit. 
Due to the combinatorial explosion of possible number of compounds as a function of atom number and types, 
this results rapidly in prohibitively large training data set needs.
ii) Transferability: Lack of generalization to new chemistries (dissimilar to training) and overfitting to known chemistries (similar to training) 
severely hampers the broad and robust applicability of ML models.
It should be stressed that random sampling of compounds found in nature introduces severe bias towards those  
atomic environments/bonding patterns that have been favored by the particular free energy reaction pathways on 
planet earth that happened to be favoured by virtue of certain boundary conditions such as 
element abundance, planetary conditions, ambient conditions and evolutionary chemistries. 
Any model is deemed useful only under the condition that decent generalizability can be achieved. }

\textcolor{black}{There also exist well-established {\em ab initio} methods to address transferability/scalability. 
For instance, order $N$ scaling methods~\cite{StechelOrderN_1994,GoedeckerOrderN_1999},
or fragmentation strategies~\cite{cr_2012_qcWithFrags} have already addressed scalability within quantum mechanics,
but typically trade accuracy or transferability for speed, or suffer from steep scaling pre-factors.
While transferable by design, popular cheminformatics models, e.g.~based on extended connectivity fingerprint descriptors
have yet to be successfully applied to quantum properties~\cite{googlePaper2017}.}

\begin{figure*}[hbtp]
\centering
\includegraphics[scale=0.8]{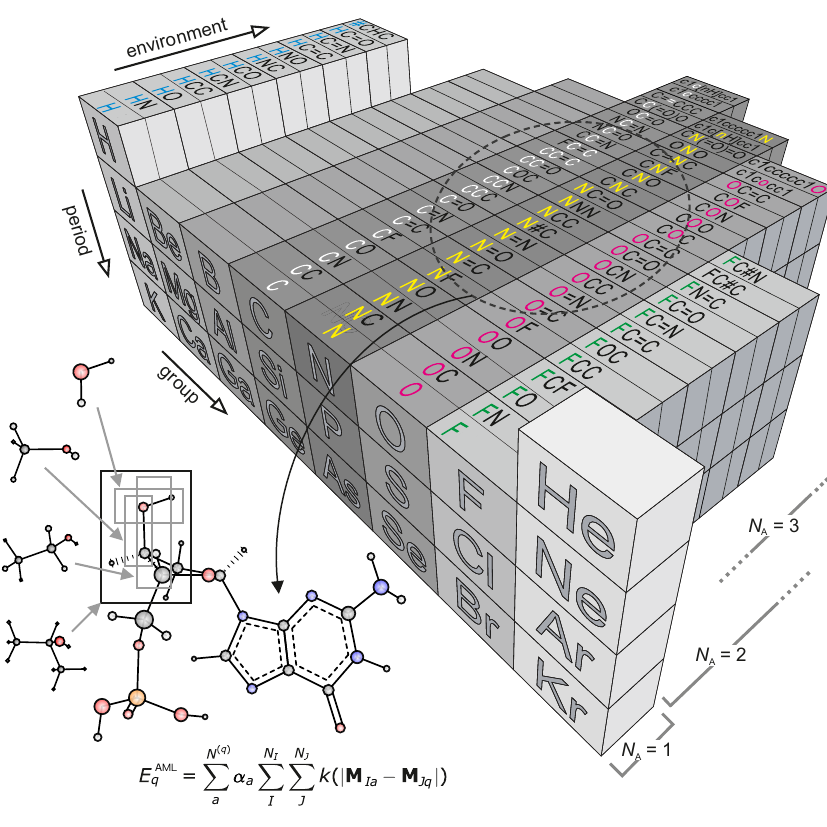} 
\caption{\label{fig:amons} 
{\bf ``amons'' ---a compositional extension of the periodic table.} 
Amon SMILES strings are arranged in increasing number of surrounding elements, representing common chemical environments. 
The amon machine learning approach estimates a property, such as the energy of a query compound $E_q$ (exemplary guanine nucleotide on display),
as an expansion in $N$ amons, the double summation being weighted by kernel ridge regression 
coefficients $\{\alpha_a\}$, and quantifying the similarity $k$ between all respective $N_J$ and $N_I$ atoms 
in query and amon molecule ${\bf M}_q$ and ${\bf M}_a$.  
} \end{figure*}

Here, we introduce a method where training instances are \textcolor{black}{generated on demand and selected} 
from a dictionary of a series of small molecular building blocks which systematically grow in size. 
\textcolor{black}{This is driven by the fact that the complete enumeration of constituting groups is feasible}, 
no matter how large and diverse the query. 
Since such building blocks repeat throughout the chemical space of larger  query compounds,
they can be viewed as indistinguishable entities, each representing an ``amon'' ({\underline A}tom-in-\underline{M}olecule-on).
In close analogy to words in dictionaries encoding the meaning of long sentences, 
amons can can encode the properties of large molecules or materials.
Explicitly accounting for all the possible chemical environments of each element, 
amons thus effectively extend the dimensionality of Mendeleev's table, as illustrated in Fig.~\ref{fig:amons}.
In this paper, we demonstrate that AML models for arbitrary query molecules and materials can be generated 
``on-the-fly'' and in a systematic fashion by training on molecules selected from a dictionary of amons 
using simple similarity measures.

\section*{Results}
\paragraph{\textbf{Amon-based ML model.}} AML models encode a Bayesian approach which infers the energy of {\em any} query compound, no matter its size
or composition, based on a linear combination of properly weighted reference results (from quantum chemistry or experiments) 
for its constituting building blocks.
The AML approach rests upon a locality assumption (known to hold under certain conditions~\cite{KohnNearsightedness,StijnPNAS2017}), 
which is exploited to systematically converge effective quantum property contributions with respect to amon size and number. 
When breaking up large query molecules, e.g.~through cascades of bond separating reactions,~\cite{Pople1970bondseparation} 
increasingly smaller and more common molecular fragments are obtained, the summed up energy of which will increasingly deviate from query.
In order to systematically control the errors resulting from our AML {\em Ansatz}, we perform the reverse procedure: 
Starting very small, increasingly larger amons (representing fragments) 
are being included in training, resulting in increasingly more accurate ML models. 
The training set size is minimized by selecting only the most relevant amons, 
i.e.~those small fragments which retain the local chemical environments encoded in the coordinates of the query molecule 
(e.g.~obtained through preceding universal force-field relaxation~\cite{MMFF94s}). 
The AML approach has limitations, e.g. when it comes to challenging non-local quantum effects, 
such as in highly electron correlated situations (Peierls/Jahn-Teller distortions, metal-insulator transitions etc).

For an efficient AML implementation, we make use of an analytical two and three-body interatomic 
potential based representation of atoms in molecules (See Method section for details). 
For training data and query validation we have used popular standard quantum chemistry protocols (see also Method section). 
For selection, a sub-graph matching procedure iterates over all non-hydrogen atoms, identifies the relevant amons, and sorts them by size ($N_I$).
This is exemplified for the organic query molecule furanylyl-propanol (C$_7$H$_{10}$O$_2$) in Fig.~\ref{sfig:algo}, 
illustrating how the amon selection algorithm dials in all the relevant local chemical environments of each atom. 

\begin{figure*}[hbtp]
\centering
\includegraphics[scale=0.7]{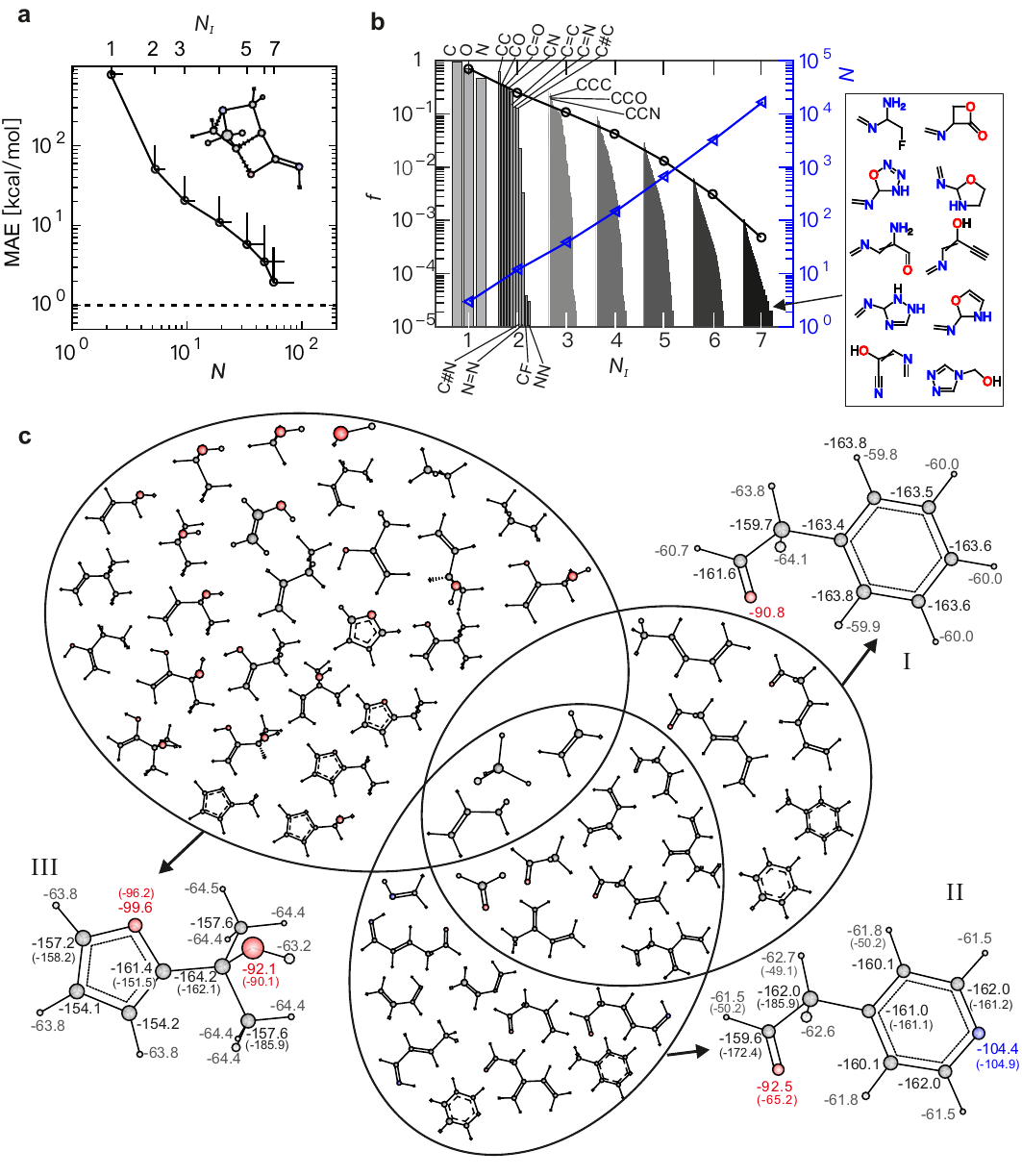}
\caption{\label{fig:gdb9} 
{\bf The amons of organic chemistry}: (A) Mean absolute prediction error (circles) of atomization energy of eleven thousand organic molecules (with $N_{I} = 9$ from QM9~\cite{gdb9_data}) 
as a function of number of heavy atoms per amon ($N_I$, not counting hydrogens) or amons ($N$) in training set. 
Standard deviations with respect to error and $N$ are also shown. 
The inset shows an outlier with high strain and prediction error of $\sim$10 kcal/mol. 
(B) Left axis: Frequency of amons ($f =$ number of occurrence/number of query molecules) in descending order; 
right axis: Number of query molecules $N$, both as a function of $N_{I}$ of amons.
Insets specify the most and least frequent amons.
(C) amons for 2-phenylacetaldehyde (\uppercase\expandafter{\romannumeral 1\relax}), 2-(furan-2-yl)propan-2-ol (\uppercase\expandafter{\romannumeral 2\relax}) and 2-(pyridin-4-yl)acetaldehyde (\uppercase\expandafter{\romannumeral 3\relax}). 
Overlapping regions correspond to shared amons.
Numbers indicate atomic energy contributions to atomization energy, regressed by AML and by Morse-potential (in brackets, see Method section for details) for those atoms where meaningful. More details about AML performance for these three molecules are available in Fig.~\ref{sfig:EandAlphaFor3QM9}.}
\end{figure*}

\paragraph{\textbf{The dictionary of organic molecules}} When exploring chemical space one invariably faces the problem of severe selection bias due to the unfathomably large scale resulting from all the possible combinations of atom types and coordinates. 
In order to rule out the possibility of AML results being coincidental, 
we have investigated the predictive performance for eleven thousand diverse organic query molecules made up of nine heavy atoms.
All query molecules were drawn at random from the QM9 dataset~\cite{gdb9_data} consisting of coordinates and electronic 
properties of 134k organic molecules from the GDB data set~\cite{gdb17}.
While GDB constitutes by no means a comprehensive subset of chemical space,
it was designed to represent important branches of chemistry~\cite{gdb17}. 
Furthermore, results of AML are obtained without loss of generality: Any other molecular data set could have been chosen just as well.
After selecting relaxed amons and subsequent training, out-of-sample prediction errors of atomization energies decrease
systematically with number and size of amons (See Fig.~\ref{fig:gdb9}A),
and nearly reach chemical accuracy ($\sim$1 kcal/mol) for average training set sizes of $\sim$50 amons with no more than seven heavy atoms. 
Corresponding standard deviations of predicted energies also decrease systematically.
By comparison, tens of thousands, and at best thousands, of training molecules are typically needed to reach similar prediction errors 
using other state of the art ML models trained on randomly selected molecules~\cite{DTNN2017,googlePaper2017,FCHL}.
Examining molecules one by one, we find that largest deviations correspond to molecules containing highly strained fragments, 
example shown as inset in Fig.~\ref{fig:gdb9}A.
In order to properly account for strained query molecules, 
amons with similarly strained local motifs would be required.

The $\sim$110k organic molecules in QM9 with nine heavy atoms can be fragmented into just $\sim$25k amons with up to seven heavy atoms (Fig.~\ref{sfig:qm9amons1k}).
While distinct, these amons are indistinguishable in the sense that they do not depend on possible query molecule, 
but can be combined to yield real-time atomization energy estimates with an expected mean absolute error of $\sim$1.6 kcal/mol.
The exponentially decaying normalized frequency distribution of amons with number of atoms is shown in Fig.~\ref{fig:gdb9}B, 
along with the exponentially growing number of possible molecules in QM9.
As one would expect, the smaller the amon the more frequently it will be selected, and for any given amon 
size high carbon content is more frequent than high oxygen or nitrogen content. 
A list and a movie, displaying the one thousand most frequent amons are provided in the supplementary materials.
Conversely, the larger the amon the less likely that it will be needed for predicting properties of a random query molecule. 
It is hence consistent that the ten least frequent amons, not shared by any pair of query molecules, 
represent rather pronounced chemical specificity (shown in Fig.~\ref{fig:gdb9}B).
These results suggest that the fundamental idea of using amon based building blocks within ML models is meaningful for chemistry:
The larger the queries and the weaker the accuracy requirements, the more amons will be shared.
As such, the AML model effectively exploits the lower dimensionality of fragments in the 
very high dimensional chemical space, known to scale exponentially with number of atoms~\cite{gdb17,anatole-ijqc2013}. When considering applications to more diverse datasets than conformer-free QM9, e.g. by including conformers or transition states, or also other chemical elements than just CHONF, the overall number of amons necessary to reach a certain threshold, will grow exponentially with the number of additional degrees of freedom. Regarding the necessary number of amons, we note that while finite in general due to the assumption of locality, this trade-off should be studied carefully before making conclusive statements on the overall efficiency and transferability of the method.

Amons manifest a deepened understanding of chemistry and are amenable to human interpretation.
For example, consider the number and nature of amons shared among different query molecules. 
They can serve as an intuitive yet rigorous measure of chemical similarity, as exemplified for three organic query molecules in Fig.~\ref{fig:gdb9}C.
The smallest amons are shared by all molecules, and the more similar a pair of query molecules the larger the overlap:
Molecules I and II are more similar than I and III, which are more similar than II and III.
Shared amons imply that AML predictions of other compounds will require fewer additional reference data.

Another valuable insight obtained from AML is the transferability of regressed atomic properties: 
Local atoms possessing similar environments will contribute to similar degrees to the total property (of extensive nature, such as energy), a feature which also underpins the atom-in-molecules theory~\cite{AIM}. 
Atomic contributions to the AML estimated atomization energies, shown for the three example molecules in Fig.~\ref{fig:gdb9}C, illustrate this point.
Note that their relative values are consistent with chemical intuition about covalent bonding. 
For example, there are three types of local oxygen environments (carbonyl, alcohol, and furane) 
which contribute to covalent binding ($\sim$92, 92, and 100 kcal/mol, respectively) in the order expected
for an atom sharing one double bond to carbon, two single bonds (carbon, hydrogen), and being in a conjugated environment (furane).
Also, $sp^3$-hybridized (with neighbors being all C/H), $sp^2$-hybridized (in carbonyl), and aromatic (in benzene) carbon atoms contribute with $\sim$160, $\sim$161,
and $\sim$164 kcal/mol, respectively, also reflecting the fact that 
aromaticity provides significant stabilization to C atoms. 
It is also consistent that hydrogen atoms have a relatively small variance in their contribution (60 to 64 kcal/mol), they can only have single bonds.
Corresponding energies from reparameterized Morse potentials~\cite{baml} compare favourably, 
suggesting the capability of AML to provide qualitative and quantitative insight to a degree 
previously only accessible through physically motivated approximations. 

\begin{figure*}[hbtp]
\centering
\includegraphics[scale=0.9]{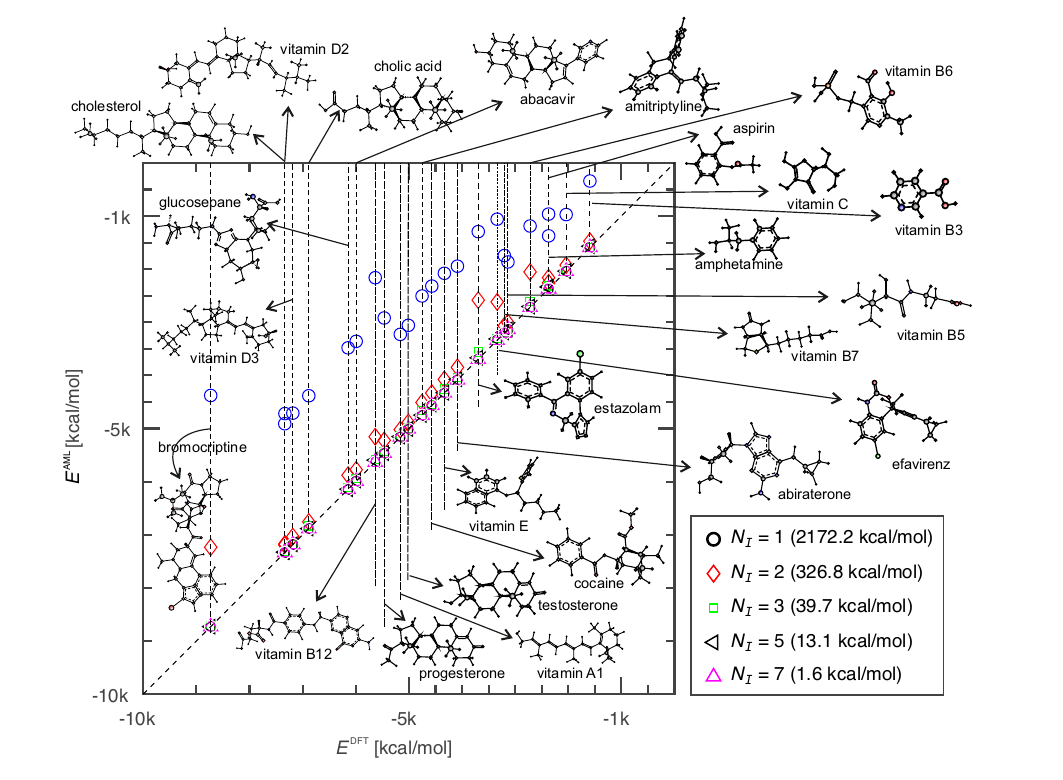} 
\caption{\label{fig:bio} 
{\bf Scalability}, demonstrated by systematic improvement of predicted atomization energies ($E$) for two dozen important biomolecules using increasingly larger amons. 
The inset specifies the maximal number of heavy atoms ($N_I$, not counting hydrogens) per amon, as well as resulting MAE. 
Chemical, DFT, and bond-counting accuracy is roughly reached for amons with 7, 5 and 3 heavy atoms, respectively. Relevant amons used are specified in Fig.~\ref{sfig:bio_amons} and supplementary data.
} 
\end{figure*}

\paragraph{\textbf{Applications of AML.}} Total energy predictions for two dozen large and important biomolecules, including cholesterol, cocaine, and vitamin D2 illustrate the 
scalability and transferability of AML. 
True versus predicted energies are shown in Fig.~\ref{fig:bio} for various AML models trained on sets of amons containing $N_\mathrm{I} =$1, 2, 3, 5, and 7 heavy atoms
(learning curves given in SI, Fig.~\ref{sfig:bio_LC}). 
Systematic improvement of predictive accuracy is found reaching errors typically 
associated with bond-counting, DFT,
or experimental thermochemistry for amons with 3, 5, or 7 heavy atoms, respectively.
For smaller query molecules with rigid and strain-less structure and homogeneous chemical 
environments of the constituting atoms, e.g., vitamin B3 with only 9 heavy atoms, 
the prediction error decreases faster with amon size than for more complex molecules, reaching 
chemical accuracy with only $\sim$20 amons in total (see Fig.~\ref{sfig:bio_LC}).
Not surprisingly, large and complex molecules with diverse atomic chemical environments, such as cholesterol, 
require substantially more amons to reach the same level of accuracy.
On the scale of atomization energies, the results also
reflect basic chemistry: Predicted energies decrease towards the reference as the amons account for contributions 
corresponding to composition ($N_I = 1$), bonds ($N_I = 2$) and angles ($N_I = 3$) between heavy atoms.

\begin{figure*}[hbtp]
\centering
\includegraphics[scale=0.75]{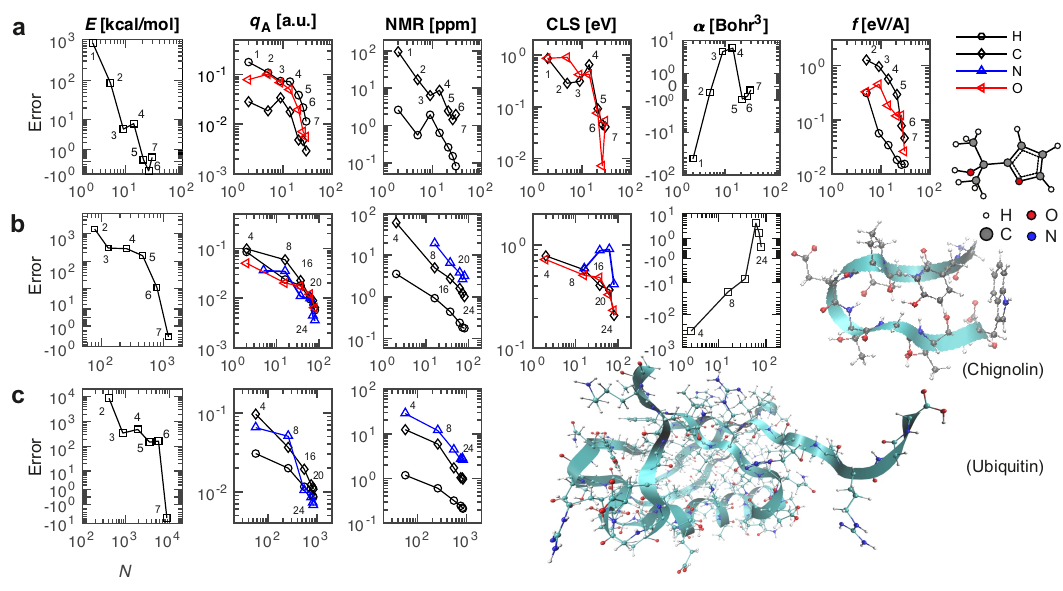}
\caption{\label{fig:multiprops} 
{\bf Applicability}: Prediction errors as function of training set size for various quantum properties 
of (A) 2-(furan-2-yl)propan-2-ol (9 heavy atoms), (B) chignolin (1UAO) (77 heavy atoms), (C) ubiquitin (1UBQ) (602 heavy atoms).
Signed prediction errors are shown for global properties only (total energy ($E$) and polarizability ($\alpha$))
with each scatter point symbolising an increase in amon size, $1 \le N_I \le 7$. 
Mean absolute errors are reported for atomic properties (charges ($q_\mathrm{A}$), NMR shifts, core level shift (CLS), force components ($f$)).
Amon sizes are specified by numbers inset. 
}
\end{figure*}

In addition to the scalability of the AML model, its general applicability to 
other quantum ground state properties, and not just energies is a direct result from 
the first principles philosophy underpinning quantum machine learning (the representation being property invariant)~\cite{QMLessayAnatole}. 
We demonstrate this fact for furanylpropanol (9 heavy atoms) by generating AML models not only for global 
properties, such as the atomization energy or isotropic polarizabilities, 
but also for atomic Mulliken charges, NMR shifts ($^1$H, $^{13}$C, $^{15}$N), core electron level shifts,
and atomic force components. Prediction geometry and training amons had to be distorted for the latter.
Identical amon training sets were used for {\em all} properties considered.
Resulting learning curves in Fig.~\ref{fig:multiprops}A indicate systematic improvement
with amon size and number, and reach or outperform the accuracy 
of hybrid DFT approximations~\cite{ChemistsGuidetoDFT}  
after training on just 30 amons (shown in Fig.~\ref{sfig:algo}) with no more than seven heavy atoms.

To demonstrate applicability {\em and} scalability, we have generated AML models 
for the deca-peptide and small hairpin model chignolin (77 heavy atoms, coordinates from PDB structure) 
as well as for the ubiquitin protein, present in all eukaryotic organisms (602 heavy atoms, coordinates from PDB structure, see supplementary data). 
Note that due to strong intra-molecular non-covalent interactions, 
unrelaxed amons were necessary, and distinct sets were required for rapid convergence of
global (energy, polarizability) and atomic (charges, NMR shifts, and core level shifts) properties, respectively. 
Because of the many electrons in these systems, a more approximate DFT functional with reduced cost and improved scalability has been used to validate AML results for these two systems
(See Method section for more details).  Figs.~\ref{fig:multiprops}B and C report corresponding learning curves (all properties but forces for chignolin,
for ubiquitin reference calculations for only energies, charges and NMR shifts were carried out). 
Again, convergence towards the reference numbers is observed, while, not surprisingly
substantially larger and more amons are required to reach small prediction errors.
The prediction error for the atomization energy decreases to less than 10 kcal/mol which is more accurate
than a standard GGA DFT for small molecules (note that DFT errors are expected to grow with molecular size).
Prediction errors for all other properties reach or exceed the accuracy of DFT~\cite{ChemistsGuidetoDFT}.
These results indicate great promise for use of AML models to contribute to NMR based protein structure determination or to first principles based molecular dynamics simulations.
In order to place these calculations into perspective, the DFT based validation calculation for ubiquitin required 
more than $\sim$4 CPU years 
and $\sim$1.6 TB of memory
using highly optimized code (see Method section) on a modern 
compute node. 
By contrast, the amon data was obtained within CPU hours on a standard laptop.
And given all the necessary amons, the AML time to solution
was CPU hours (minutes) for the
computation of atomic kernel matrix
for the atomization energy (charges and NMR shifts).

To further assess the applicability of AML models, we have investigated five different classes of systems,
each representing important yet different chemistries. 
Regardless of size and chemical nature, AML prediction errors decrease systematically and 
reach rapidly chemical accuracy as number and size of selected amons grow: 
(i) Five industrially relevant polymers of increasing size and chemical complexity have been considered including
polyethylene ($N_{I} =$ 26), polyacetylene ($N_{I} =$ 30), alanine peptide ($N_{I} =$ 50), polylactic acid ($N_{I} =$ 50) 
and the backbone of quaternary ammonium polysulphone ($N_{I} =$ 96),
the latter being essential for alkaline polymer electrolyte fuel cells~\cite{QAPS_PNAS}.
Resulting learning curves in Fig.~\ref{sfig:sca}A reveal the familiar trend: 
The more chemically complex the system, the more amons are necessary to achieve the chemical accuracy threshold of $\sim$1 kcal/mol:
Polysulphone, the most complex polymer out of the five, requires nearly 75 times more amons ($\sim$300) 
than polyethylene, the chemically simplest polymer ($N_a\sim10$). 
(ii) Prediction errors of AML models of water clusters with up to 21 molecules~\cite{h2oN_TIP5P} 
decay rapidly and systematically for all clusters, and reach chemical accuracy with at most seven amons 
and no more than ten water molecules/amons, providing additional evidence for the applicability to non-covalent bonding 
(see Fig.~\ref{sfig:sca}B).
(iii) Hexagonal BN sheets doped with carbon and gold (C-$h$BN; Au-$h$BN; C,Au-$h$BN) have previously been
reported to efficiently catalyze CO oxidation~\cite{Au_hBN}. 
Due to the underlying symmetries in such ordered materials, 
AML model prediction errors converge to chemical accuracy within at most eleven amons 
for the most complex C,Au-doped $h$BN variant (see Fig.~\ref{sfig:sca}C). 
(iv) Symmetry plays an even more important role in crystalline bulk: 
To accurately predict the cohesive energy of silicon only three amons with no more than eighteen atoms are required (Fig.~\ref{sfig:sca}D). 
(v) Finally, we have also considered one of the most important non-covalent bonding patterns
in nature, the Watson-Crick base pairing in DNA, Cytosine-Guanine and Adenine-Thymine.
While amons with no more than seven heavy atoms suffice to reach chemical accuracy for individual base pairs, 
amons corresponding to truncated motifs of hydrogen bonds with up to eleven heavy atoms are necessary to properly account
for the hydrogen bonding (see Fig.~\ref{sfig:CG} and Fig.~\ref{sfig:AT}).
It is not surprising that non-covalent bonding patterns, such as Watson-Crick bonding, require larger amons, 
as they extend over larger spatial domains. 
Furthermore, they also exhibit strong non-local effects through their conjugated moieties, implying the need 
for amons with multiple hydrogen bonds (see Fig.~\ref{sfig:CG} and Fig.~\ref{sfig:AT} for amons that contain
simultaneously aromatic fragments and hydrogen bonds) to achieve chemical accuracy.

\begin{figure}[h]
\begin{center}
\includegraphics[scale=0.2, angle=0]{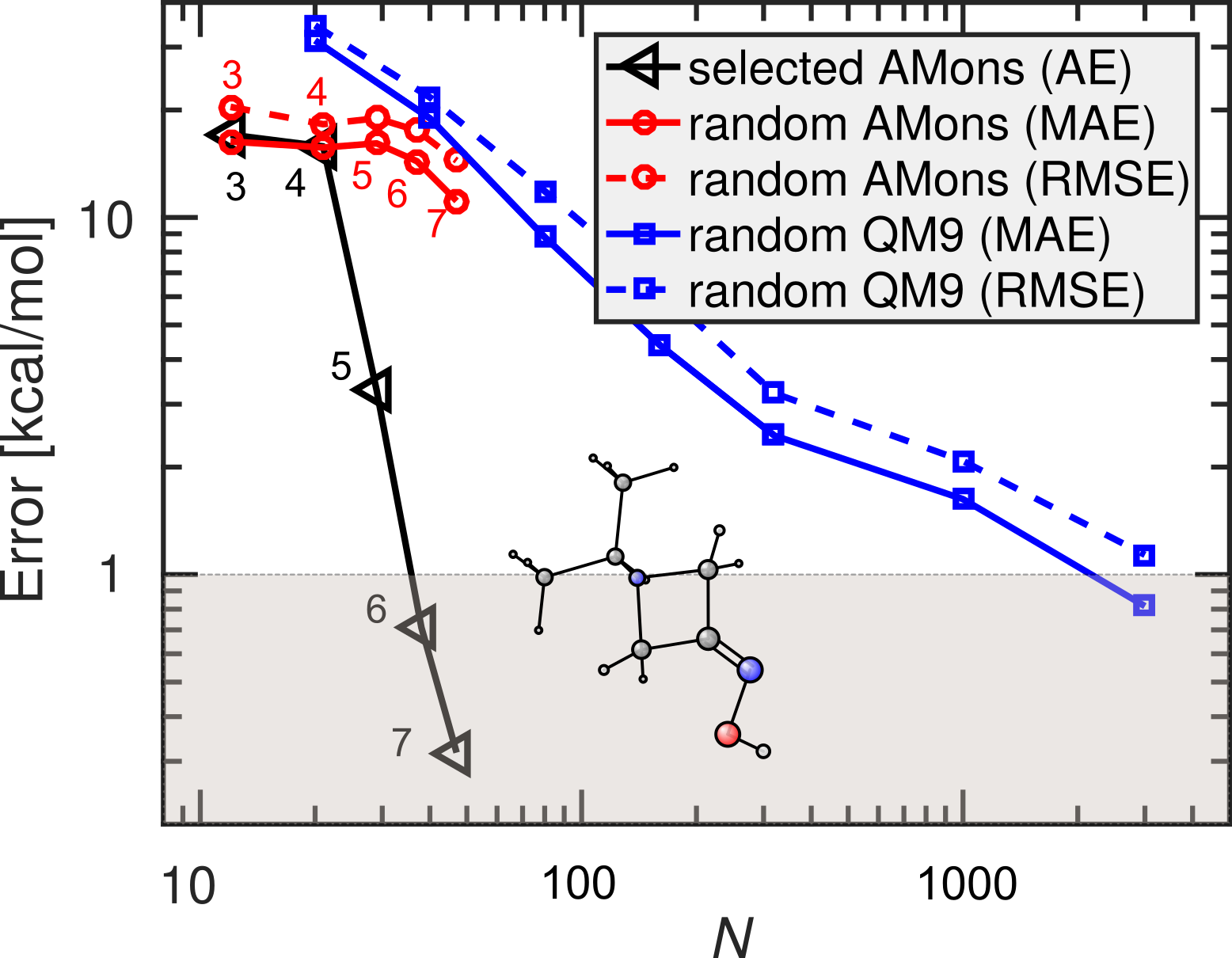}
\label{fig:Klaus}
\caption{
ML prediction error of atomization energy of molecule shown in inset as a function of training set size. 
Using amon based selection chemical accuracy domain (gray, $\le$ 1 kcal/mol) is reached with tens of small training molecules (black),
rather than thousands of large molecules (black) drawn at random from QM9.
Black: Absolute error (AE) of AML model (using selected amons).
Red: Mean absolute error (MAE) and root mean square error (RMSE) of one hundred AML models (using shuffled random amons of QM9 molecules,
obtained previously for generating AML learning curves in Fig. 2 A). 
black: MAEs and RMSEs of one hundred conventional ML model directly trained on randomly 
drawn QM9 molecules. 
Red and black numbers indicate respective amon-sizes for AML models.
}
\end{center}
\end{figure}

\textcolor{black}{
So far, we have discussed performance and results which promise to {\bf simultaneously} reaching 
unprecedented levels of data efficiency, accuracy, scalability, and transferability. 
The ``on-the-fly-selection'' of relevant amons results in considerable 
boosts to the learning rate, as can be seen in Fig. 2 A for predictions of
$\sim$11'000 organic molecules. They reach 2 kcal/mol with less than $\sim$50 selected training amons.
By comparison, such accuracies are reached with conventional (non-scalable) methods only after training on thousands of training molecules.
In order to better contextualize the boost obtained, Fig.~5 displays 
a direct comparison for a typical strain-free QM9 molecule.
It is obvious that, in order to reach chemical accuracy, the AML approach requires only tens of small training 
instances (no more than 7 atoms), rather than thousands of large training molecules (up to 9 atoms) in the case of random training set selection.
}

\textcolor{black}{
Finally, we have investigated more specifically the limitations arising due to the locality assumptions for the case of delocalized electronic states.
More specifically, Fig.~\ref{sfig:conj} details prediction errors 
as a function of length of poly-acetylene as a query system. 
As one would expect, the error increases with query length. 
It is obvious, however, that the larger the amons in the training set, the lower the off-set of the error curve.
In particular, for amons containing no more than 3 acetylene units ($N_I =$ 6), the prediction error remains below the
chemical accuracy threshold of 1 kcal/mol for query systems with up to 14 carbon atoms. 
Increasing that number to 4 or 5 acetylene units affords chemical accuracy for predicting systems with up to 28 or more than 30 carbon atoms, respectively.
This, in combination with the use of periodic amons (e.g. see $h$BN sheets, Fig.~\ref{sfig:sca}C) for infinite systems,
clearly demonstrates that also delocalized systems can be dealt with in a systematic fashion which exceeds the state of the art in ML.
For comparison, the error of a linear-scaling \emph{ab initio} method for poly-acetylene with $N_I=56$ is also shown~\cite{MTA_PA_2010}. 
While substantially lower than AML predictions with 5 units, 
the method in question can only reduce the total computational time to a half of that required by a traditional DFT. 
}


\section*{Conclusion}
To conclude, our numerical evidence suggests that a finite set of small to medium sized
building block molecules (dubbed ``amons'') is sufficient for the automatized 
generation of quantum machine learning models with favorable combination of 
efficiency, accuracy, scalability, and transferability. 
Given an amon dictionary, such models can be used to rapidly estimate ground state quantum properties of large and real materials with chemical accuracy. 
Consequently, we think it evident that atomistic simulation protocols reaching the accuracy of
high-level reference methods, such as post-Hartree-Fock or quantum Monte Carlo, for large systems
are no longer prohibitive in the foreseeable future, 
effectively sidestepping the various accuracy issues which have plagued many common DFT approximations~\cite{Perdew2017Science}.
Other future work will deal with amon dictionary extensions to  more degrees of freedom and more chemical elements. 
Eventually, open-shell, reactive, charged and electronically excited species can be considered.



\clearpage




\section*{Methods}


\paragraph{\textbf{Amon selection}}

This section describes the amon selection algorithm in detail (see Fig.~\ref{sfig:algo}).
Given a query molecule with nuclear charges \{$Z$\} and coordinates \{$R$\}, we first establish its connectivity table through the use of covalent radii ($r_{\mathrm{cov}}$) taken from reference~\cite{rvdw}. Namely, we consider two atoms to be covalently bonded if their distance is less or equal to the sum of their respective covalent radii plus 0.5, meanwhile maintaining the maximal connectivity of C/N atoms to be no more than 4/3 (this criteria is essential for correctly determining connectivity of strained molecule). Afterwards, we assign bond orders (1, 2 or 3) to bonds based on the connectivity table of each atom in the molecule.

Next, the connectivity table, together with element symbols and coordinates are stored to a SDF file, and then loaded by the OEChem toolkit~\cite{oechem}. 
Essentially, now we have perceived, from molecular geometry, the molecular graph with vertices \{$V_0$\} and edges \{$E_0$\} being weighted respectively by the atom types (i.e., nuclear charge) and bond orders.
Note that i) this step would be skipped if we were given as input SMILES strings; ii) hydrogens are being neglected throughout the entire amon selection procedure, as their numbers and positions can be determined once bond orders and positions of heavy atoms are known. 
	
Before proceeding, one important step is to identify all the hybridization states of all heavy atoms (excluding hydrogens), especially those multivalent atomic environments, including S atom with 6 valences (i.e., R-S(=O)(=O)-R', where R and R' are some functional groups) and N atom with 5 valences (i.e., R-N(=O)=O), vital for valence consistency check. 
Therewith we build a set of connected (i.e., there exists a path between any two vertexes in the graph) subgraphs \{$g_i$\} from the query molecular graph $G_0$ = \{$V_0$, $E_0$\} and loop through each of the thus-obtained subgraphs. 
For any subgraph, if any of its vertices (i.e., atoms) does not preserve the original hybridization state, it's not considered as a valid amon. This implies that we cannot break any double bond in query molecule to generate fragments as training instances. 
This is especially notable for molecules containing the aforementioned multivalent atoms. For example, suppose the query molecule is RS(=O)(=O)R', by breaking one of the S=O bonds, we end up with a fragment R[SH]=O, which is not a valid amon as there is a significant change of valence from 6 in query molecule to 4 now for sulfur atom. 
The same argument holds true for other bonds with bond order larger or equal to 2, e.g., C=C, C\#N, etc.

The next step is to check whether or not a subgraph $g_i = \{v_i, e_i\}$ is isomorphic to any subgraph of the query molecular graph $G_0$. This is true if and only if there exists a \emph{one-to-one} mapping between each vertex of $\{v_i\}$ and a vertex in \{$V_0$\}, and between each vertex in $\{v_i\}$ and some edge in \{$E_0$\}. So to be isomorphic, one has an exact match. Obviously, the edge counts for each vertex is maintained.
As a note in passing, another similar concept, subgraph monomorphism, may cause confusion. A subgraph $g_i$ is monomorphic to a subgraph of $G_0$ (subgraph monomorphism) if and only if there is a mapping between vertices of $g_i$ and $G_0$, and if there is also an edge between all vertices in $\{v_i\}$ for which there is a corresponding edge between all vertices in $\{V_0\}$. Apparently, edge counts in this case may vary.
By imposing subgraph isomorphism instead of subgraph monomorphism on amons, the local environments can be retained in a more faithful manner, meanwhile reducing the number of possible graphs to some extent. Very often, subgraph mismatch for amons that are subgraph monomorphic are found for ring systems, such as the five-membered furane system in Fig.~\ref{fig:gdb9} of the main text, where C=COC=C is a connected monomorphic subgraph with all local hybridization states retained, but does not meet the criteria of subgraph isomorphism, in contrast to C1=COC=C1. Therefore, the latter is selected as a valid amon as it undoubtedly recovers the aromatic local environment in the query molecule, while the former is discarded. 

Filtered subgraphs thus-far are not yet guaranteed to be representative fragments. First, an additional geometry relaxation is performed for the corresponding fragment (now with valencies saturated with hydrogen atoms) using MMFF94s force field (FF)~\cite{MMFF94s} as implemented in Openbabel~\cite{obabel}, and with dihedral angles fixed to match the local geometry of the query molecule (to avoid significant conformational change in local environment). This step is followed by a DFT relaxation using a quantum chemistry code (we use Gaussian 09~\cite{Gaussian09}),
if the resulting amon candidate has not already been calculated previously for some other query molecule.    
At this stage it can happen that the amon candidate dissociates (turning into a non-connected graph), or that it reacts chemically to transform into a molecule with a different connectivity. In the former case, the fragment should be discarded, in the latter case, the subgraph isomorphism has to be confirmed again. 
We proceed with the amon candidate if the fragment has experienced no change in connectivity, or if subgraph isomorphism is retained despite geometry changes. 
If the resulting fragment has not been seen before, i.e., the root mean squared distance between the geometries of this fragment and any other fragment accepted so far is larger than zero, we add it to the amon pool. 

As the number of atoms in amon candidates is being increased, one continues looping through subgraphs until they have been exhausted. The resulting amon pool is considered as the ``optimal'' training set for the query molecule.
The above procedure has been implemented in the AQML code~\cite{aqmlcode} using either SMILES string or DFT coordinates as input. Note that query coordinates of other form (e.g., that resulting from less expensive force-field methods, such as~\cite{MMFF94s}) can be used just as well~\cite{DeltaPaper2015}.
Figs.~\ref{sfig:bio_amons}-\ref{sfig:polymers_amons} show amons used to predict energies of amphetamine, aspirin and cholesterol, QM9 dataset (only the most frequent 1k) and five common polymers. All coordinates as well as energies for those amons not shown are included in this contribution as xyz files in supplementary data.
	
For systems involving significant van der Walls (vdW) interaction, such as water clusters and the Watson-Crick DNA base pair Guanine-Cytosine, the only change that needs to be made to the above algorithm is to assign a separate set of bonds
for the pair of atoms with bond order smaller than 1, say 0.5, based on the summation of vdW radius.
More specifically, if the distance between two non-covalently bonded atoms A and B is less or equal to the summation of vdW radius of A and B scaled by 1.25, then the associated two heavy atoms are considered connected through a vdW bond.
Resulting amons are thus termed vdW amons.
As a note, hydrogen bonds can be also naturally covered due to its short-ranged (relatively speaking) nature than other types of non-covalent bonds, thus we are able to treat these two types of interactions in a unified fashion. 
The vdW radius used here are taken from reference~\cite{rvdw}.

There are cases that long-ranged interaction has a remarkable effect on some atomic property (e.g., NMR shifts) of atoms in complex systems, such as protein.
The aforementioned algorithm, however, can only account for property for which short-ranged interactions dominate, such as energy.
To incorporate long-ranged effects in amons, we therefore need for a new strategy. Currently, we simply assign amons as the complete and unique cutouts of query molecule, 
i.e., any amon is made up of an heavy atom in the target plus its local fragment (saturated by hydrogen atoms) 
enclosed within a cutoff radii of 3.6 Angstrom centered on that atom (totalling no more than
24 heavy atoms). 
Two complex systems considered in this research (chignolin and ubiquitin, see more details in the subsection Computational details) and corresponding amons included as xyz files in supplementary data.
    
In principle, one would always optimize the geometry of amons to gain transferability throughout chemical space; 
however, due to i) geometry relaxation for large amons may be expensive (in particular for accurate methods like CCSD(T)), while a single point energy calculation is computationally much cheaper and 
ii) there is no need for a dictionary for one or several systems chosen as proof-of-concept, we determined to use instead an alternative set of amons, 
for each of which the coordinates of heavy atoms (and hydrogens whenever present in the query molecule) are exactly the same as in the corresponding query molecule. 
This particular set of amons is thus dubbed ``static'' amons to distinguish from those with relaxed geometries. 
Static amons were used to test on two systems: chignolin and ubiquitin, for which vdW interactions are significant.
Efforts into reaping fully the transferability of small vdW amons ($N_I\le 7$) with relaxed geometry constitues part of future work.
    
For water clusters, the algorithm above is not applicable in practise because the local motifs of larger water clusters are usually metastable, 
resulting in substantial geometry changes after relaxation. 
For water clusters we therefore simply consider the presence of a ring structure (with edges being the vdW bond assigned to two oxygen atoms, as elucidated above for vdW amons) as the only criterion to determine if a fragment is a valid amon for a larger query cluster. 
More specifically, given a query water cluster containing only 4- and 5-membered rings, any smaller water cluster made up of only 4 or 5-membered ring structures would be considered a valid amon. 
Also, water cluster consisting of just 1 or 2 water molecules shall always be included in the amons set as their existence is universal in any larger water cluster.
The final amons chosen and larger query water clusters are displayed in Fig.~\ref{sfig:water}A and B, together with correpsonding learning curves shown in Fig.~\ref{sfig:water}C and D. 

Regarding amons selection of condensed phase systems, no general algorithm is available for now due to the non-trivial nature of symmetry consideration in amons selection algorithm.
For these systems one can simply fragment the query and select those fragments that are most representative of local environments of the query.
For elemental silicon, in spite of the fact that there is only one atom per primitive unit cell, its amons have to include increasingly larger shells of neighboring atoms, saturated with hydrogens.
To conform with the high symmetry of Si bulk, only very few high-symmetry amons are necessary to converge the energy prediction to below chemical accuracy (see Fig.~\ref{sfig:sca}D). 
For doped $h$BN sheets with a substantially larger unit cell than in Si, smaller unit cells containing the dopant are selected as amons (see Fig.~\ref{sfig:hBN}).

\paragraph{\textbf{Computational details.}}

The majority of geometries and associated energies were calculated at the level of theory B3LYP/cc-pVDZ using ORCA 4~\cite{orca4}. 
Exceptions are:
1) water clusters, for which methods like HF, MP2 and CCSD(T) were also considered. Geometries were optimized at the MP2/6-31G* level by Gaussian 09~\cite{Gaussian09} 
and the thus-obtained geometries were used to obtain single-point CCSD(T) energies using Molpro~\cite{MOLPRO} and the same Pople basis set; 
2) systems with periodic boundary conditions ($h$BN sheets, bulk silicon and their corresponding amons), for which VASP~\cite{Kresse_VASP} was employed with exchange-correlation interaction described by PBE~\cite{PBE1996} functional. 
Wavefunction was described by projected augmented wave (PAW) method~\cite{Blochl_PAW} with a plane-wave cutoff of 340 eV. 

Special attention should be paid to the geometry optimization of molecular amons:
to account for the local geometries in query molecule as much as possible, 
it is reasonable for the structure of any amon to be optimized to a local minima, by means of specifying a lower criteria for geometry optimization (see README files in supplementary data for details). 
Otherwise, the amon molecule could be optimized to a very different conformer including new local environments such as hydrogen bonds, 
which may not be present in the query molecule and thus cannot be served as one of the optimal training instances.
	
For chignolin (PDB code name: 1UAO) and ubiquitin (PDB code name: 1UBQ), experimental geometries (retrieved from online protein databank, www.rcsb.org/structure/) were 
pre-processed before carrying out DFT calculations, i.e., charge neutralization and removal of any solvent. 
For 1UAO, single point energy, polarizability, atomic charges, NMR shifts were calculated at the B3LYP level with basis cc-pVDZ by ORCA 4~\cite{orca4}. 
For the calculation of core level shifts of heavy atoms (excluding hydrogens), 
VASP was employed with the same set of computational parameters as for water clusters described above 
and the so-called initial state approximation was used. 
For 1UBQ, consisting of more than one thousand atoms, an affordable model PBE/Def2-SVP plus D3 correction was chosen and calculations were done by Turbomole~\cite{tmole}. 
The same method was used to calculate NMR shifts and atomic charges. 
For both 1UAO and 1UBQ, \emph{single point} calculations for the corresponding amons were conducted at the same level as for query molecule (i.e., no geometry optimization was carried out)
so as to fully capture the local interactions involved in query molecules. 
Note that there is overwhelming difference in the computational time for the big protein 1UBQ and all of its amons,
i.e., on the order of 10$^5$ and 10$^1$ cpu hours, respectively. 
As a comparison, for the ML part regarding atomic properties, the computational time is negligible.

For greater details on reproducing the reference data, please refer to the README files in supplementary data.
	
\paragraph{\textbf{Gaussian process (GP) regression.}}

	We start from total energy partitioning into atomic contributions. Within the atom-in-molecule (AIM)~\cite{AIM} theory 
	the total energy is an exact sum of atomic energies,
	\beq \label{eq:aim}
	E = \sum_I E^I = \sum_I \int_{\Omega_{I}} \langle \Psi | \hat{H } | \Psi \rangle d^3r
	\eeq 
	where $\Omega_I$ is the atomic basin determined by the zero-flux condition of 1-electron charge density.  
	Apparently, each atomic energy includes all short-ranged covalent bonding as well as long-ranged non-covalent bonding (e.g., van der Waals interaction, Coulomb interaction, etc.). 
	As such, similar local chemical environments of an atom in a molecule always lead to similar atomic energies~\cite{anatole-ijqc2013,aCM_NMR_Anatole}.
	Accordingly, we assume the following Bayesian model~\cite{GP} of atomic energies, 
	\beq \label{eq:bayesian_atom}
	E^I - \varepsilon_0^I = \phi(\mathbf{M}^I)^{\top}\mathbf{w} + \epsilon 
	\eeq 
	where $\epsilon$ is the error, $\mathbf{M}^I$ is an atomic representation of atom $I$ in the molecule and $\varepsilon_0^I$ is the corresponding atom-type dependent offset, which is on the scale of the energy of a free atom. 
	Without this atom-type specific shift, the distribution of atomic energies would be multimodal.
	By summing up terms at both sides in equation~\ref{eq:bayesian_atom}, we have 
	\bea
	E - \sum_I \varepsilon_0^I = \sum_I \phi(\mathbf{M}^I)^{\top}\mathbf{w} + \epsilon ' 
	\eea
	and following Bart\'ok et al.~\cite{GAP} the covariance of two molecules can be written as
	\bea
	K_{ij} = \mathrm{Cov}(E_i, E_j) = \mathrm{Cov}(\sum_I E_i^I, \sum_J E_j^J) = \sum_I \sum_J \mathrm{Cov}(E_i^I,E_j^J) = \sum_I \sum_J K_{ij}^{IJ}
	\eea
	where $I$ and $J$ run over all the atom indices in molecule $i$ and $j$, respectively. 
	The global covariance matrix element is expressed as a summation of local atomic covariance matrix elements, 
	which can be written as a function of their atomic representations:
	\beq \label{eq:covIJ}
	K^{IJ}_{ij} = k(\mathbf{M}^I_i, \mathbf{M}^J_j) = \delta_{t_I t_J} \exp \left( -\frac{| \mathbf{M}^I_i -\mathbf{M}^J_j |^2}{2\sigma_{IJ}^2} \right)
	\eeq 
	where $t_I$ is the type of atom $I$ (which can be simply characterized by nuclear charge), $\delta_{t_I t_J}$ is the Kronecker delta, $\mathbf{M}^I_i$ is the representation of atom $I$ in molecule $i$, $\sigma_{IJ}$ is the width of Gaussian kernel (other kernel functions could have been used just as well) and $L_2$ norm (i.e., Euclidean distance) is used to measure the distance between two atomic environments. Note that $\delta_{t_I t_J}$ in the above equation may be removed when some ``alchemical'' representation is used, capable of interpolating efficiently between very different atoms such as carbon and fluorine.
		
	Combined with Gaussian process regression~\cite{GP}, we arrive at the formula for the energy prediction of query molecule $q$ out-of-sample,
	\bea \label{eq:workhorse}
	E_q = \sum_i K_{qi}\alpha_i 
	\eea 
	where $\alpha_i = \sum_{j} ([\mathbf{K} + \lambda^2\mathbf{I}]^{-1})_{ij} E_j$ is the regression coefficient for the $i$-th molecule, with $\lambda$ being the regularization parameter. Rephrasing the above equation yields atomic energy of $J$-th atom in molecule $r$ ($E_q^J$), i.e., 
	\beq
	E_q = \sum_{J} E_q^J = \sum_{J} \sum_i \alpha_i\sum_{I} K_{iq}^{IJ}
	\eeq 
	
	Except for energies, the above regression procedure is equally well applicable to other size-extensive properties such as isotropic static molecular polarizability, through substitution of the energy in equation~\ref{eq:workhorse} by polarizability, leaving the kernel matrix untouched. 

\paragraph{\textbf{The SLATM representation.}}
		
	Following Ref~\cite{baml}, we chose the same starting point as in Ref.~\cite{OAvL_FRD}, i.e.~the charge density of the system. 
	\beq 
	\rho(\mathbf{r}) = \sum_i \rho^{(i)}(\mathbf{r})
	\eeq
	where $\rho^{(i)}$ is the charge density of electron $i$. For ML, we do not need accurate $\rho^{(i)}$ as an input; instead, a very coarse charge density suffices, as long as it is capable of completely capturing all the essential features of the system, i.e., geometry $\{R_I\}$ and composition $\{Z_I\}$. 
	To further simplify the problem, we consider the charge density distribution of an ensemble of electrons partitioned onto different atoms
	\beq 
	\rho(\mathbf{r}) = \sum_I \rho^I(\mathbf{r})
	\eeq 
	where we approximate $\rho^I$ as
	\beq \label{eq:charge_exapansion}
	\rho^I(\mathbf{r}) = Z_I \left ( \delta(\mathbf{r}-\mathbf{R}_I) +  \frac{1}{2}\sum_{J\neq I} Z_J\delta(\mathbf{r}-\mathbf{R}_{J})g(\mathbf{r}-\mathbf{R}_I) \right )
	\eeq 
	where $\delta(\cdot)$ is set to a normalized Gaussian function $\delta(x) = \frac{1}{ \sigma \sqrt{2\pi} }e^{ - \frac{x^2}{2\sigma} }$,
	$g(r)$ is some distance dependent scaling function, capturing the locality of chemical bonds and chosen in a form similar to the London potential $h(R) = \frac{1}{R^6}$, which is  much better than the Coulomb potential for describing covalently bonded systems with kernel ridge regression based ML models, as was demonstrated with numerical evidence in reference~\cite{baml}. 
	A factor of 1/2 before the summation term removes double counting and reflects the assumption that the amount of charge density contribution for the $I$-th atom from the $J$-th atom is the same as that for the $J$-th atom from the $I$-th atom. 
	
	The atomic ensemble charge density is still translation and rotation dependent, which is redundant for most molecular properties and introduces unnecessary degrees of freedom. To remove these redundency, we could either project $\rho^I$ to a set of basis function as was done in SOAP~\cite{soap_2013}. Instead, we represent the ensemble charge density within an internal coordinate system by projecting $\rho^I(r)$ to different internal degree of freedoms (similar to Ref.~\cite{OAvL_FRD}), associated with well-known many-body terms. That is, for the one-body term, the projection results in a scalar, i.e., the nuclear charge $Z_I$; for the two-body term, we use
	\beq \label{eq:bop}
	\frac{1}{2} \sum_{J\neq I} Z_J\delta(\mathbf{r}-\mathbf{R}_{IJ})g(\mathbf{r})
	\eeq 
	For 3-body term, we use
	\beq \label{eq:bot}
	\frac{1}{3} \sum_{J\neq K \neq I} Z_J Z_K \delta(\theta - \theta_{IJK})h(\theta, \mathbf{R}_{IJ}, \mathbf{R}_{IK}) 
	\eeq
	where $\theta$ is the angle spanned by vector $\mathbf{R}_{IJ}$ and $\mathbf{R}_{IK}$ (i.e.,$\theta_{IJK}$) and treated as a continuous variable). 
	$h(\theta, \mathbf{R}_{IJ}, \mathbf{R}_{IK})$ is the 3-body contribution depending on both internuclear distances and angle, and is chosen to correspond to the Axilrod-Teller-Muto (ATM)~\cite{atm,atm2} van der Waals potential
	\bea 
	h(\theta, \mathbf{R}_{IJ}, \mathbf{R}_{IK}) &=& \frac{1+\cos \theta \cos \theta_{JKI}\cos \theta_{KIJ}}{(R_{IJ}R_{IK}R_{KJ})^3} \\
	\theta_{JKI} = f_1(\theta, R_{IJ}, R_{IK}),
	\theta_{KIJ} &=& f_2(\theta, R_{IJ}, R_{IK}),
	R_{KJ} = f_3(\theta, R_{IJ}, R_{IK})
	\eea
	which guarantees an even faster decay than the 2-body London term.
	
	Now we can either consider $\rho^I(\mathbf{r})$ as a function of all the possible types of atoms ($\{Z_I\}$), pairs of atoms (i.e., bonds formed between $Z_I$ and $Z_J$, not necessarily covalent), triples of atoms, i.e.,
	\beq 
	\rho^I(\mathbf{r}) = \rho^I(\{t_{p}\}, \{t_{pq}\}, \{t_{pqr}\})
	\eeq 
	or alternatively in an alchemical~\cite{anatole-ijqc2013} way,
	\beq 
	\rho^I(\mathbf{r}) = \rho^I(Z, R, \theta)
	\eeq 
	Sticking to the former, the charge ensemble representation is in essence the concatenation of different many-body potential spectra, i.e., 
	\beq 
	\mathbf{M}^I = [ Z_I, \{\rho_I^{IJ}(R)\}, \{\rho_I^{IJK}(\theta)\} ],~J\neq I, K\neq J \neq I
	\eeq 
	where $\rho_I^{IJ}(R)$ (two-body term) and $\rho_I^{IJK}(\theta)$ (three-body term) are essentially radial distributions of London potential and angular distribution of ATM potentials, respectively. The resulting representation is dubbed atomic Spectrum of London and Axilrod-Teller-Muto potential (aSLATM) (see Fig.~\ref{sfig:SLATM1} for the graphical illustration of SLATM for one exemplified molecule in Fig.~\ref{fig:gdb9}A in the main text). 
	
	The distance between any two local atomic environments is then calculated as the $L_2$ norm (i.e., Euclidean distance) between their corresponding $\mathbf{M}$'s, comparing only terms sharing the same many-body type, i.e.,
	\beq 
	d(I,J) = d(\mathbf{M}^I, \mathbf{M}^J) = \sqrt[]{d_1^2 + \sum_{KL} d_2(\rho_I^{IK},\rho_J^{JL})^2 + \sum_{KMLN} d_3(\rho_I^{IKM},\rho_J^{JLN})^2 }
	\eeq 
	where $K,L,M,N$ are atomic indices, $d_1$ is the $L_2$ distance between the one-body terms of atoms $I$ and $J$,
	\beq 
	d_1(I,J) = \sqrt {Z_{I}^2 + s(I,J)Z_J^2}
	\eeq 
	with $s()$ being the sign function, 
	\beq 
	s(I,J) = \left\{\begin{matrix}
		-1,~Z_I=Z_J\\ 
		1,~Z_I\ne Z_J
	\end{matrix}\right.
	\eeq 
	$d_2$ is the $L_2$ distance between the two-body terms of atom $I$ and $J$ (truncated at an inter-atomic distance of $r_c$),
	\beq 
	d_2(\rho_I^{IK}, \rho_J^{JL}) = \left\{\begin{matrix}
		\sqrt{\int_0^{r_c} (\rho_I^{IK}(R)-\rho_J^{JL}(R))^2 dR},~Z_I=Z_J~\mathrm{and}~Z_K=Z_L\\ 
		0,~\mathrm{otherwise} 
	\end{matrix}\right.
	\eeq 
	and similarly $d_3$ characterizes the similarity between the three-body terms of the two atoms,
	\beq 
	d_3(\rho_I^{IKM}, \rho_J^{JLN}) = \left\{\begin{matrix}
		\sqrt{\int_0^{\pi} (\rho_I^{IKM}(\theta)-\rho_J^{JLN}(\theta))^2 dR},~Z_I=Z_J,~Z_K=Z_L~\mathrm{and}~Z_M=Z_N\\ 
		0,~\mathrm{otherwise}. 
	\end{matrix}\right.
	\eeq 
	By binning each many-body term, a representation vector is formed for each atom in a molecule. By concatenating these atomic representations, we arrive at the so-called local representation of a molecule. A even simpler variant is the so-called global representation, which is, in our case, a summation of those atomic vectors.
	The corresponding size of global or each atomic representation is fixed for any dataset built out of a given set of elements. 
	
Note that the above approach for constructing representation, may remind one of the widely acknowledged many-body expansion (MBE) approach for resolving the dispersion interaction in rare-gas molecules in physics. Indeed, the expression of $n$-body terms ($n=2,3$) used here are almost identical as in the MBE approach; however, they are fundamentally different when it comes to representation, that is, the kernel can account for interaction of much higher order even using a representation of lower order in nature, effectively circumventing the issue of convergence~\cite{MBEconvergeIssue} using a truncated series in MBE to directly calculate the interaction strength.

	Regarding the generation of both SLATM and aSLATM representations, 
	three parameters are involved: 1) Cutoff radii $r_c$ for the 2-body term. 
	Since London and ATM potential guarantees extremely fast decay of atomic 
	interactions, a cutoff radii of 4.8 {\AA} is sufficient; 
	2) Width $\sigma$ of the Gaussian ``smearing'' function. For radial and terms, $\sigma$ is set to 0.05 {\AA}. While for angular terms, 0.05 rad is used. 
	3) Grid spacing $h$. In practise, (a)SLATM is calculated on a set of radial and angular grids and thus $h$ determines the computational cost. A large $h$ is computationally efficient, but the accuracy is likely to be a problem; while for a very small $h$, it would be too expensive to generate the representation. Convergence tests show that when $h$ is set to 0.03 {\AA} (rad) for the radial (angular) term, a balance between accuracy and efficiency is achieved.

	SLATM and aSLATM have been tested for global ML models trained on randomly selected molecules, 
	in analogy to the assessment in Ref.~\cite{googlePaper2017}. 
	In order to compare the relative performance between ML models based on
	SLATM and other representations proposed in literature, three datasets: QM7b~\cite{CM}, QM9~\cite{gdb9_data,gdb17} and 6k isomers (a subset of QM9) were considered. For all these datasets, random sampling was used to generate training set, and the remaining was selected for test. For QM9 dataset~\cite{gdb9_data}, 229 molecules out of 133885 molecules 
	dissociated after optimization, and were not considered for this study.
	The corresponding indices of these 229 dissociated molecules along with their input SMILES string are given in the supplementary information of \cite{baml}.
	Fig.~\ref{sfig:SLATM2} illustrate their predictive power by comparison to results obtained using the Coulomb matrix (CM)~\cite{CM}, Bag of Bond (BoB)~\cite{BoB}, and BAML~\cite{baml} representations.

	All errors reported refer to out-of-sample test set deviations from reference, meaning
	that these samples had no part in the training of the ML models. 
	For KRR-ML models based on discretized representations of molecules (or atoms) such as CM, BoB and BAML, the Laplacian kernel was found to be better than Gaussian kernel; while the inverse is true for continuous representations such as SLATM. 
	In both cases, hyperparameters (including regularization parameter $\lambda$ and kernel width $\sigma$) were chosen by default schemes. 
	Specifically, four settings are used respectively for four different scenarios:
	
	i) for Laplacian kernel plus global representation (could be one of CM/BoB/BAML), $\lambda = 1\times 10^{-10}$, $\sigma = \max (\mathbf{D})/c$, where $\mathbf{D}$ is the distance matrix (based on the representation being used) between pairs of molecules, $c$ is a scaling factor equal to $\ln(2)/5$
	
	ii) for Gaussian kernel + global SLATM, $\lambda = 1\times10^{-4}$, $c = \sqrt{2\ln(2)}/5$.

For AML models, aSLATM is always used together with Gaussian kernel. Hyperparameters differ for the other two cases:

	iii) for AML of extensive properties (energy, polarizability), $\lambda = 1\times10^{-4}$, $c = \sqrt{2\ln(2)}$ for energy and $c=2.5 * \sqrt{2\ln(2)}$ for polarizability.
	
	iv) for AML of atomic properties (atomic charge, NMR shifts, core level shifts), $\lambda = 1\times10^{-8}$, $c = \sqrt{2\ln(2)}$.

	Part of these heuristic choices of kernel widths are inspired by Ref~\cite{one_kernel_Raghu}.

\paragraph{\textbf{Morse-potential based atomic energies}}
	We truncate the many-body expansion of the total energy at 2-body terms, which are approximated by Morse potential with relevant parameters retrieved from the supplementary material of reference~\cite{baml}. 
	In order to obtain atomic contributions to the atomization energy ($E_A$), 
	we follow the same strategy as adopted in Mulliken charge population, i.e., the atomic contribution of atom $I$ to the total $E_A$ is simply the summation of all bonds involving atom $I$ divided by 2. 
	The thus-obtained atomic energies are more transferable than bond energies (which typically depends on bond order), as indicated by the values within brackets in Fig.~\ref{fig:gdb9}C.

\section*{Data availability}
All data used in this paper are available at {\tt https://github.com/binghuang2018/aqml-data}. 
All pertinent details are specified in the {\tt README} file.

\section*{Code availability}

Mixed Python/Fortran code (MIT license, no restrictions) for generating amons, aSLATM/SLATM representation as
well as AML models are available, along with detailed instructions on how to reproduce 
our results, at {\tt https://github.com/binghuang2018/aqml}.




\begin{thebibliography}{10}

\bibitem{feynman1963}
R.~P. Feynman, R.~B. Leighton, M.~Sands, {\it The Feynman lectures on physics.
  Vol. 1\/} (Addison-Wesley, 1963).

\bibitem{martin2004}
R.~M. Martin, {\it Electronic structure: basic theory and practical methods\/}
  (Cambridge university press, 2004).

\bibitem{biology2011}
J.~B. Reece, {\it et~al.\/}, {\it Campbell biology\/} (Pearson Boston, 2011).

\bibitem{CM}
M.~Rupp, A.~Tkatchenko, K.-R. M\"uller, O.~A. von Lilienfeld, {\it Phys. Rev.
  Lett.\/} {\bf 108}, 058301 (2012).

\bibitem{ML4Polymers_Rampi2013}
G.~Pilania, C.~Wang, X.~Jiang, S.~Rajasekaran, R.~Ramprasad, {\it Scientific
  reports\/} {\bf 3}, 2810 (2013).

\bibitem{ML4Crystals_Wolverton2014}
B.~Meredig, {\it et~al.\/}, {\it Phys. Rev. B\/} {\bf 89}, 094104 (2014).

\bibitem{Alan_OLED2015}
E.~O. Pyzer-Knapp, K.~Li, A.~Aspuru-Guzik, {\it Adv. Fun. Mat.\/} {\bf 25},
  6495 (2015).

\bibitem{Felix2016}
F.~A. Faber, A.~Lindmaa, O.~A. von Lilienfeld, R.~Armiento, {\it Phys. Rev.
  Lett.\/} {\bf 117}, 135502 (2016).

\bibitem{Sandip2016}
S.~De, A.~P. Bartok, G.~Csanyi, M.~Ceriotti, {\it Phys. Chem. Chem. Phys.\/}
  {\bf 18}, 13754 (2016).

\bibitem{DTNN2017}
K.~T. Sch{\"u}tt, F.~Arbabzadah, S.~Chmiela, K.~R. M{\"u}ller, A.~Tkatchenko,
  {\it Nat. Commun.\/} {\bf 8}, 13890 (2017).

\bibitem{GrossmannCNN2018}
T.~Xie, J.~C. Grossman, {\it Phys. Rev. Lett.\/} {\bf 120}, 145301 (2018).

\bibitem{ANI}
J. S. Smith, O. Isayev and  A. E. Roitberg, {\it Chem. Sci.\/} {\bf 8}, 3192 (2017).

\bibitem{SchNet}
K. T. Sch{\"u}tt, H. E. Sauceda, P.-J. Kindermans, A. Tkatchenko and K.-R. M{\"u}ller, {\it J. Chem. Phys.\/} {\bf 148}, 241722 (2018).

\bibitem{Shapeev_ActiveLearning}
K. Gubaev, E. V. Podryabinkin and A. V. Shapeev, {\it J. Chem. Phys.\/} {\bf 148}, 241727 (2018).

\bibitem{Ceriotti_AutoKernelSelect}
G. Imbalzano, A. Anelli, D. Giofr{\'e}, S. Klees, J. Behler and M. Ceriotti, {\it J. Chem. Phys.\/} {\bf 148}, 241727 (2018).

\bibitem{cr_2012_qcWithFrags}
M. S. Gordon, D. G. Fedorov‡Spencer, R. Pruitt and L. V. Slipchenko, {\it Chem. Rev.\/} {\bf 112}, 632 (2012).


\bibitem{KohnNearsightedness}
E.~Prodan, W.~Kohn, {\it Proc. Natl. Acad. Sci. USA\/} {\bf 102}, 11635 (2005).

\bibitem{StijnPNAS2017}
S.~Fias, F.~Heidar-Zadeh, P.~Geerlings, P.~W. Ayers, {\it Proc. 
Natl. Acad. Sci. USA\/} {\bf 114}, 11633 (2017).

\bibitem{Pople1970bondseparation}
W.~J. Hehre, R.~Ditchfield, L.~Radom, J.~A. Pople, {\it J. Am. Chem. Soc.\/}
  {\bf 92}, 4796 (1970).

\bibitem{MMFF94s}
T.~A. Halgren, {\it J. Comp. Chem.\/} {\bf 20}, 720 (1999).

\bibitem{gdb9_data}
R.~Ramakrishnan, P.~Dral, M.~Rupp, O.~A. von Lilienfeld, {\it Sci. Data\/} {\bf
  1}, 140022 (2014).

\bibitem{gdb17}
L.~Ruddigkeit, R.~van Deursen, L.~C. Blum, J.-L. Reymond, {\it J. Chem. Inf.
  Model.\/} {\bf 52}, 2864 (2012).

\bibitem{googlePaper2017}
F.~A. Faber, {\it et~al.\/}, {\it J. Chem. Theory Comput.\/} {\bf 13}, 5255
  (2017).

\bibitem{FCHL}
F.~A. Faber, A.~S. Christensen, B.~Huang, O.~A. von Lilienfeld, {\it J. Chem. Phys.\/} {\bf 148}, 241717 (2018).

\bibitem{anatole-ijqc2013}
O.~A. von Lilienfeld, {\it Int. J. Quantum Chem.\/} {\bf 113}, 1676 (2013).

\bibitem{AIM}
R.~F. Bader, {\it Atoms in molecules\/} (Wiley Online Library, 1990).

\bibitem{baml}
B.~Huang, O.~A. von Lilienfeld, {\it J. Chem. Phys.\/} {\bf 145}, 161102
  (2016).

\bibitem{QMLessayAnatole}
O.~A. von Lilienfeld, {\it Angew. Chem. Int. Ed.\/} {\bf 57}, 4164 (2018).

\bibitem{ChemistsGuidetoDFT}
W.~Koch, M.~C. Holthausen, {\it A Chemist's Guide to Density Functional
  Theory\/} (Wiley-VCH, 2002).

\bibitem{QAPS_PNAS}
S.~Lu, J.~Pan, A.~Huang, L.~Zhuang, J.~Lu, {\it Proc. Natl. Acad. Sci. USA\/}
  {\bf 105}, 20611 (2008).

\bibitem{h2oN_TIP5P}
T.~James, D.~J. Wales, J.~Hernández-Rojas, {\it Chem. Phys. Lett.\/} {\bf
  415}, 302 (2005).

\bibitem{Au_hBN}
K.~Mao, {\it et~al.\/}, {\it Sci. Rep.\/} {\bf 4}, 5441 (2014).

\bibitem{Perdew2017Science}
M.~G. Medvedev, I.~S. Bushmarinov, J.~Sun, J.~P. Perdew, K.~A. Lyssenko, {\it
  Science\/} {\bf 355}, 49 (2017).

\bibitem{oechem}
Oechem toolkit 2.1.2, openeye scientific software (2017).

\bibitem{obabel}
N.~M. O'Boyle, {\it et~al.\/}, {\it J. Cheminform.\/} {\bf 3}, 1 (2011).

\bibitem{Gaussian09}
M.~J. Frisch, {\it et~al.\/}, Gaussian~09 {R}evision {D}.01. Gaussian Inc.
  Wallingford CT 2009.

\bibitem{aqmlcode}
B.~Huang, A.~von~Lilienfeld, AQML: Amons-based quantum machine learning code for quantum chemistry, https://github.com/binghuang2018/aqml (2020).
  
\bibitem{DeltaPaper2015}
R.~Ramakrishnan, P.~Dral, M.~Rupp, O.~A. von Lilienfeld, {\it J. Chem. Theory
  Comput.\/} {\bf 11}, 2087 (2015).

\bibitem{rvdw}
M.~Mantina, A.~C. Chamberlin, R.~Valero, C.~J. Cramer, D.~G. Truhlar, {\it The
  Journal of Physical Chemistry A\/} {\bf 113}, 5806 (2009).

\bibitem{MOLPRO}
H.-J. Werner, {\it et~al.\/}, Molpro, version 2015.1, a package of ab initio
  programs (2015).

\bibitem{Kresse_VASP}
G.~Kresse, J.~Furthmüller, {\it Comp. Mat. Sci.\/} {\bf 6}, 15 (1996).

\bibitem{PBE1996}
J.~P. Perdew, K.~Burke, M.~Ernzerhof, {\it Phys. Rev. Lett.\/} {\bf 77}, 3865
  (1996).

\bibitem{Blochl_PAW}
P.~E. Bl\"ochl, {\it Phys. Rev. B\/} {\bf 50}, 17953 (1994).

\bibitem{tmole}
{TURBOMOLE V6.2 2010}, a development of {University of Karlsruhe} and
  {Forschungszentrum Karlsruhe GmbH}, 1989-2007, {TURBOMOLE GmbH}, since 2007;
  available from \\ {\tt http://www.turbomole.com}.

\bibitem{pbeh3c}
S.~Grimme, J.~G. Brandenburg, C.~Bannwarth, A.~Hansen, {\it J.
  Chem. Phys.\/} {\bf 143}, 054107 (2015).

\bibitem{orca4}
F.~Neese, {\it WIREs: Comput Molecul Sci.\/} {\bf 2}, 73 (2012).

\bibitem{aCM_NMR_Anatole}
M.~Rupp, R.~Ramakrishnan, O.~A. von Lilienfeld, {\it J. Phys. Chem. Lett.\/}
  {\bf 6}, 3309 (2015).

\bibitem{GP}
C.~Rasmussen, C.~Williams, {\it Gaussian Processes for Machine Learning\/},
  Adaptative computation and machine learning series (University Press Group
  Limited, 2006).

\bibitem{GAP}
A.~P. Bart\'ok, M.~C. Payne, R.~Kondor, G.~Cs\'anyi, {\it Phys. Rev. Lett.\/}
  {\bf 104}, 136403 (2010).

\bibitem{one_kernel_Raghu}
R.~Ramakrishnan, O.~A. von Lilienfeld, {\it CHIMIA\/} {\bf 69}, 182 (2015).

\bibitem{OAvL_FRD}
O.~A. von Lilienfeld, R.~Ramakrishnan, M.~Rupp, A.~Knoll, {\it Int. J. Quantum
  Chem.\/} {\bf 115}, 1084 (2015).

\bibitem{soap_2013}
A.~P. Bart\'ok, R.~Kondor, G.~Cs\'anyi, {\it Phys. Rev. B\/} {\bf 87}, 184115
  (2013).

\bibitem{atm}
B.~M. Axilrod, E.~Teller, {\it J. Chem. Phys.\/} {\bf 11}, 299 (1943).

\bibitem{atm2}
Y.~Muto, {\it J. Phys.-Math. Soc. Japan.\/} {\bf 17}, 629 (1943).

\bibitem{MBEconvergeIssue}
M.~Doran, I.~Zucker, {\it J. Phys. C: Sol. Stat. Phys.\/} {\bf 4},
  307 (1971).

\bibitem{BoB}
K.~Hansen, F.~Biegler, O.~A. von Lilienfeld, K.-R. M\"uller, A.~Tkatchenko,
  {\it J. Phys. Chem. Lett.\/} {\bf 6}, 2326 (2015).

\bibitem{Sibulk_a0}
Y.~Okada, Y.~Tokumaru, {\it J. Appl. Phys.\/} {\bf 56}, 314 (1984).

\bibitem{Sibulk_Ecoh}
B.~Farid, R.~Godby, {\it Phys. Rev. B\/} {\bf 43}, 14248 (1991).

\bibitem{MTA_PA_2010} 
S.~D.~Yeole, S.~R.~Gadre, {\it The Journal of Chemical Physics}, {\bf 132}, 094102 (2010).

\bibitem{StechelOrderN_1994}
W. Hierse and E. B. Stechel, {\it Phys. Rev. B}, {\bf 50}, 17811 (1994).

\bibitem{GoedeckerOrderN_1999}
S. Goedecker, {\it Rev. Mod. Phys.}, {\bf 71}, 1085 (1999)

\end{thebibliography}


\section*{Acknowledgements}
D. Bakowies is acknowledged for helpful discussions. O.A.v.L. acknowledges funding from the Swiss National Science foundation (No.~PP00P2\_138932 and 407540\_167186 NFP 75 Big Data). 
This research was partly supported by the NCCR MARVEL, funded by the Swiss National Science Foundation.
Calculations were performed at sciCORE (http://scicore.unibas.ch/) scientific computing core facility at University of Basel. 

\section*{Author contributions}
All authors contributed extensively to the work presented in this paper.

\section*{Additional information}
Supplementary Information accompanies this paper at http://www.nature.com. 


\clearpage

\section*{Supplementary Information}

\subsection*{Supplementary Movie}

A movie named \texttt{Amons1k.mov} is provided, displaying the 1k most frequent amons being used by QM9 dataset.

\setcounter{figure}{0}
\renewcommand{\thefigure}{S\arabic{figure}}

\subsection*{Supplementary Figures}

	\begin{figure}[!htb]
		\centering
		\includegraphics[scale=0.6]{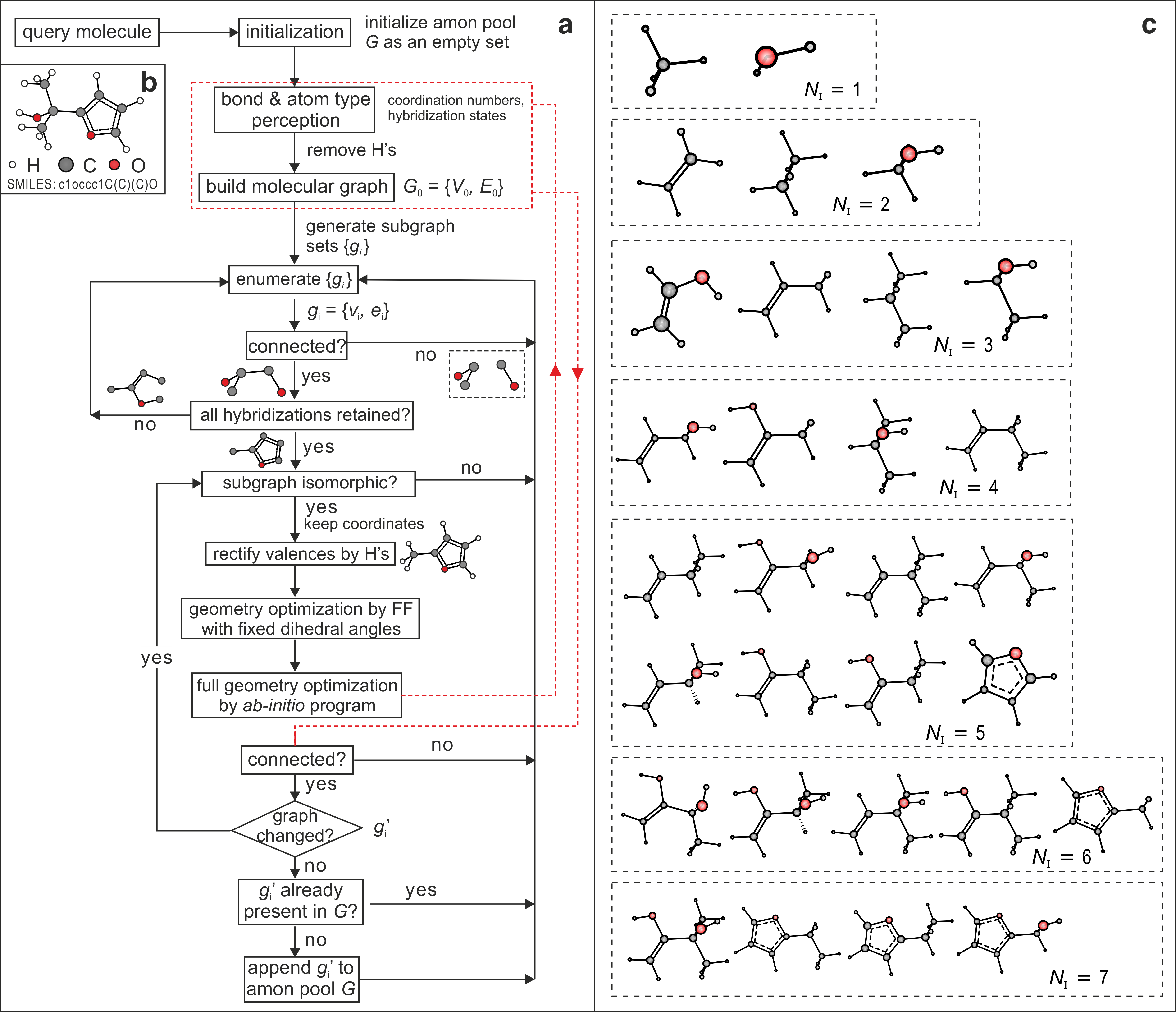}
		\caption{\label{sfig:algo}
			Detailed flowchart for amon selection algorithm (A), as exemplified for a query molecule 2-(furan-2-yl)propan-2-ol (B). The selected amons for the query are displayed in (C).}
	\end{figure}
	\clearpage

\begin{figure} 
		\centering
		\includegraphics[scale=0.45]{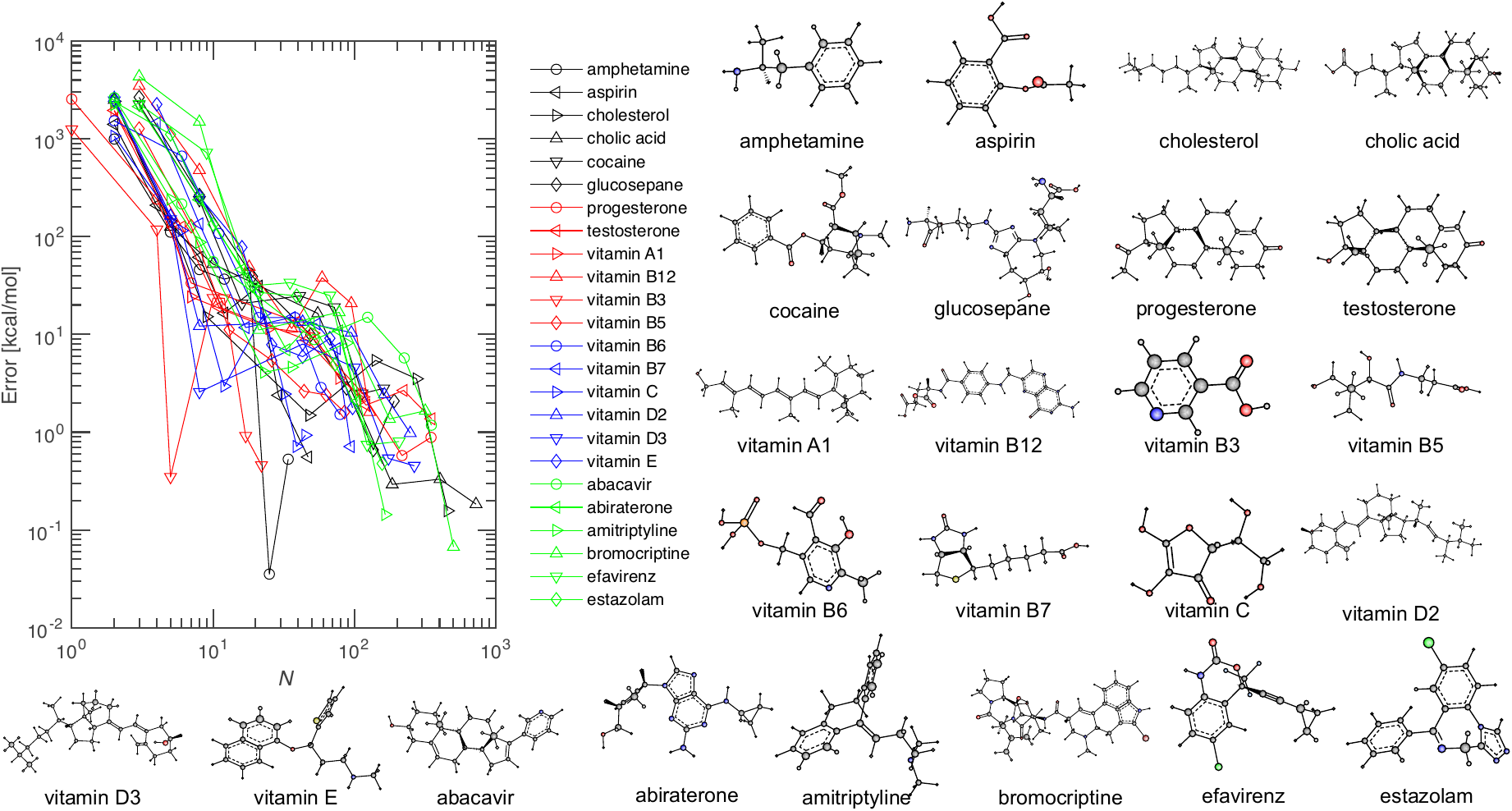}
         \caption{\label{sfig:bio_LC}Absolute prediction error as a function of number of amons used for training for 24 query biomolecules (also shown in Fig.~\ref{fig:bio} of the main text).}
  	\end{figure}
  	\clearpage



\begin{figure*} 
\centering
\includegraphics[scale=0.7]{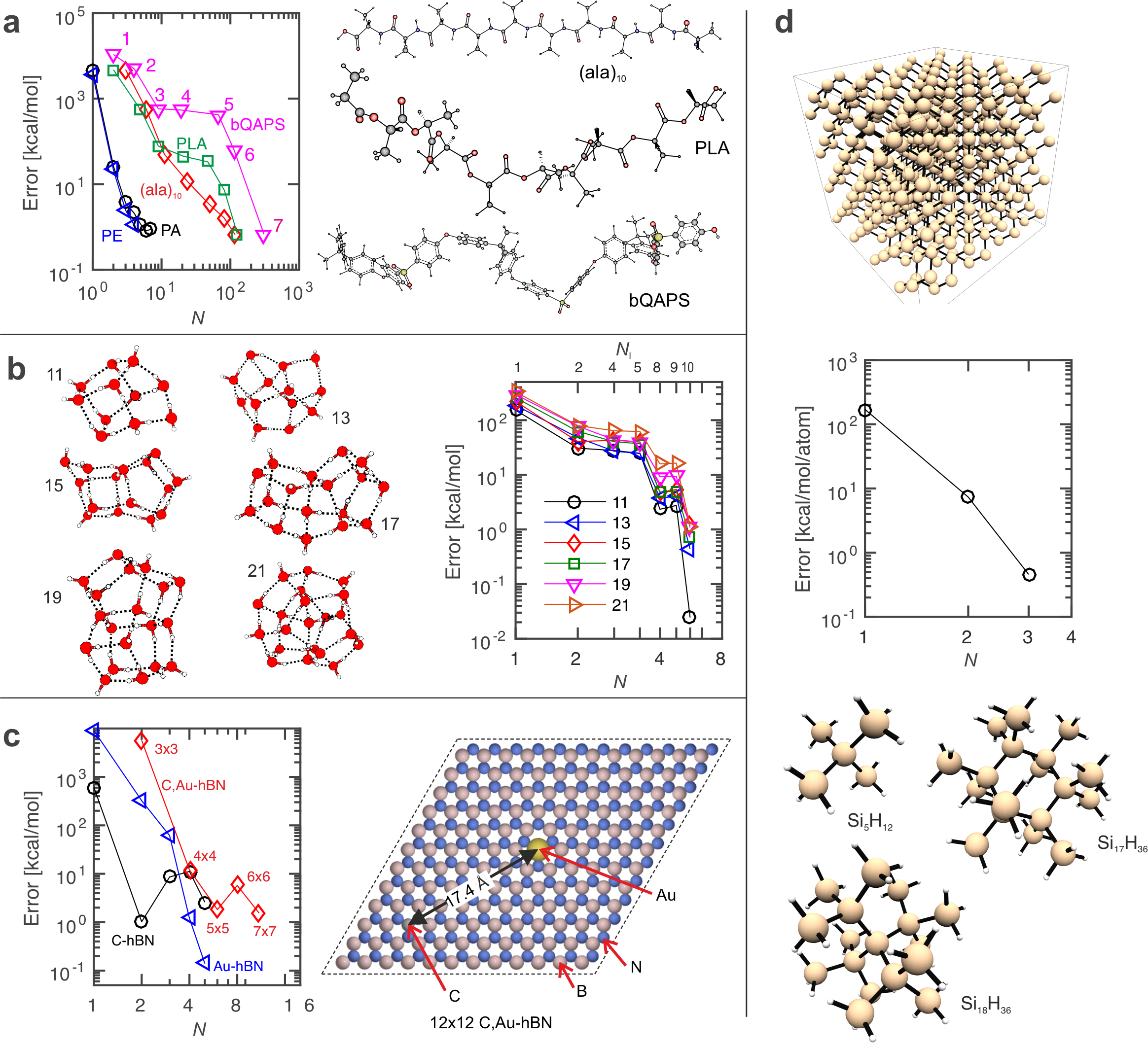}
\caption{\label{sfig:sca} 
{\bf Scalability of the AML model}, illustrated by learning curves for (A) 6 polymers, including polyethylene (PE, with 28 monomers), polyacetylene (PA, with 15 monomers), alanine peptide ((ala)$_{10}$), 
polylactic acid (PLA, with 10 monomers) 
and the backbone of quaternary ammonium polysulphone (bQAPS, with 3 monomers);
(B) 6 water clusters with number of water molecules being 11, 13, 15, 17, 19, and 21;
(C) 2-D hexagonal BN sheets with one B atom replaced by gold (Au-$h$BN), or carbon (C-$h$BN), or two B atoms replaced by one gold and carbon (Au,C-$h$BN).
Absolute errors are shown for per unit cell (each unit cell is of size 12x12). See Fig.~\ref{sfig:hBN} for corresponding amons.
(D) Bulk silicon. Amons are displayed in the bottom panel, with chemical formula attached to each amon.
Inset integers indicate the size of amons (i.e., number of heavy atoms).}
\end{figure*}


\begin{figure*}
\centering 
\includegraphics[scale=0.7]{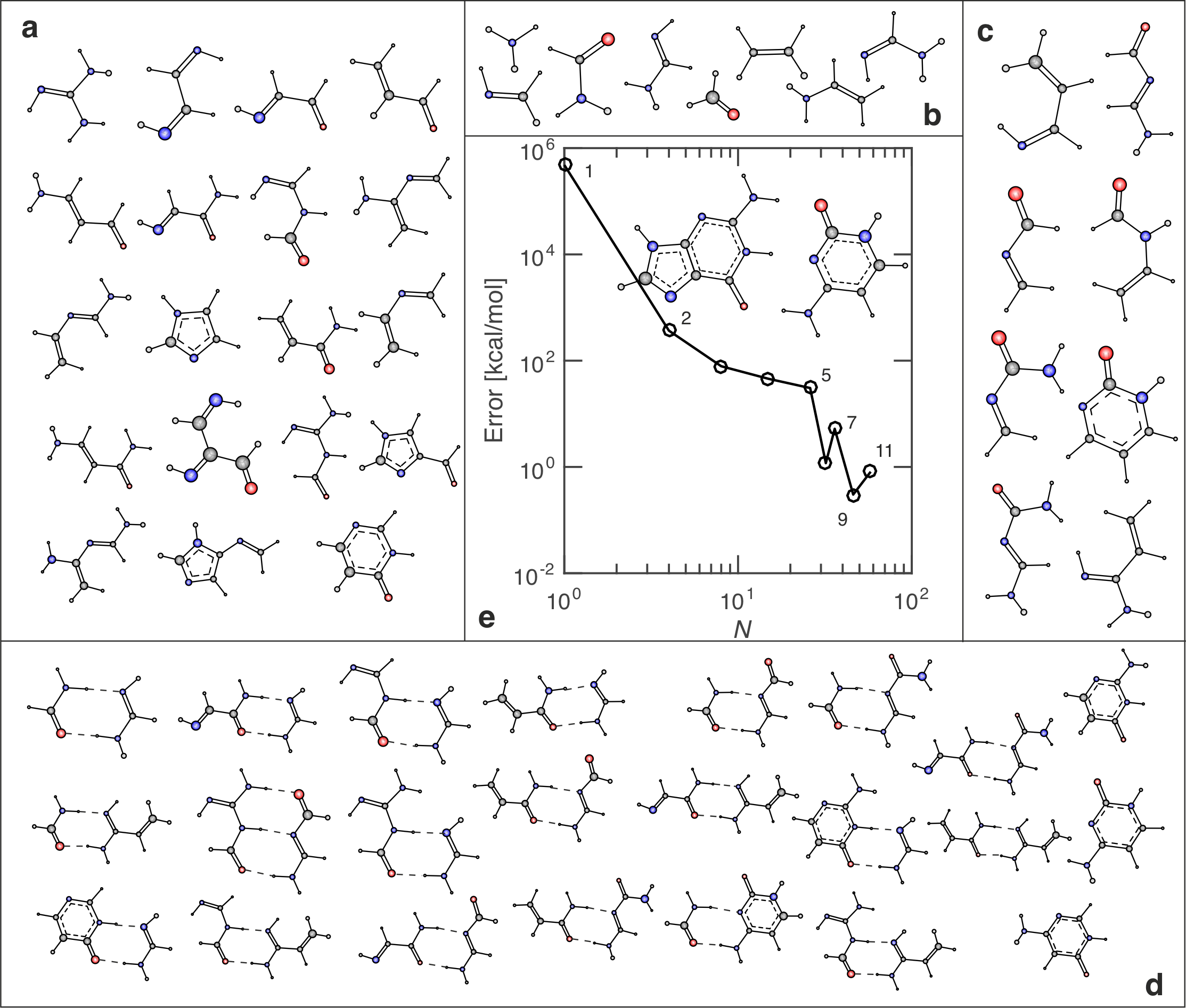}
\caption{\label{sfig:CG} 
{\bf The amons of DNA}, illustrated for Watson-Crick base-pair GC (inset of E). (A) amons of DNA base Guanine;
(B) Amons shared by Cytosine and Guanine;
(C) Amons of Cytosine;
(D) Amons of Watson-Crick bonding pattern, all complexes containing at least two hydrogen-bonds. 
(E) Learning curve for total energy prediction of Cytosine-Guanine (CG) base pair. Trained on all amons in A-D, AML underestimates the DFT energy of GC by 0.81 kcal/mol.}
\end{figure*}

\begin{figure*}
\centering 
\includegraphics[scale=0.9]{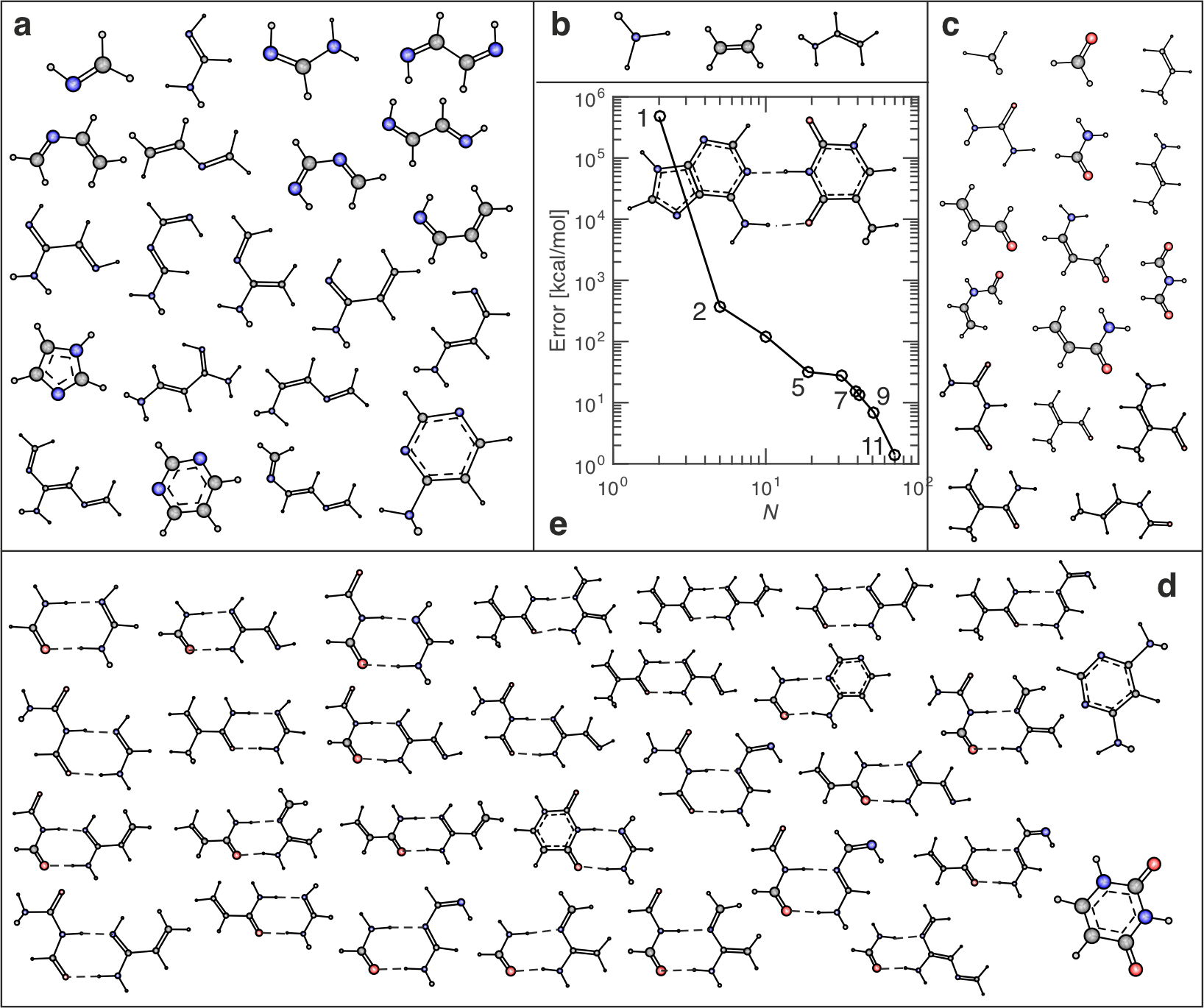}
\caption{\label{sfig:AT} 
{\bf The amons of DNA}, illustrated for Watson-Crick base-pair AT (inset of E).
(A) Amons of DNA base Adenine;
(B) Amons shared by Adenine and Thymine;
(C) Amons of Thymine;
(D) Amons of Watson-Crick bonding pattern, all complexes containing at least two hydrogen-bonds.
(E) Learning curve for the total energy prediction of Adenine-Thymine (AT) base pair. Trained on all amons in A-D, AML underestimates the DFT energy of AT by 1.41 kcal/mol.}
\end{figure*}

	\begin{figure}
		\centering
		\includegraphics[scale=0.6]{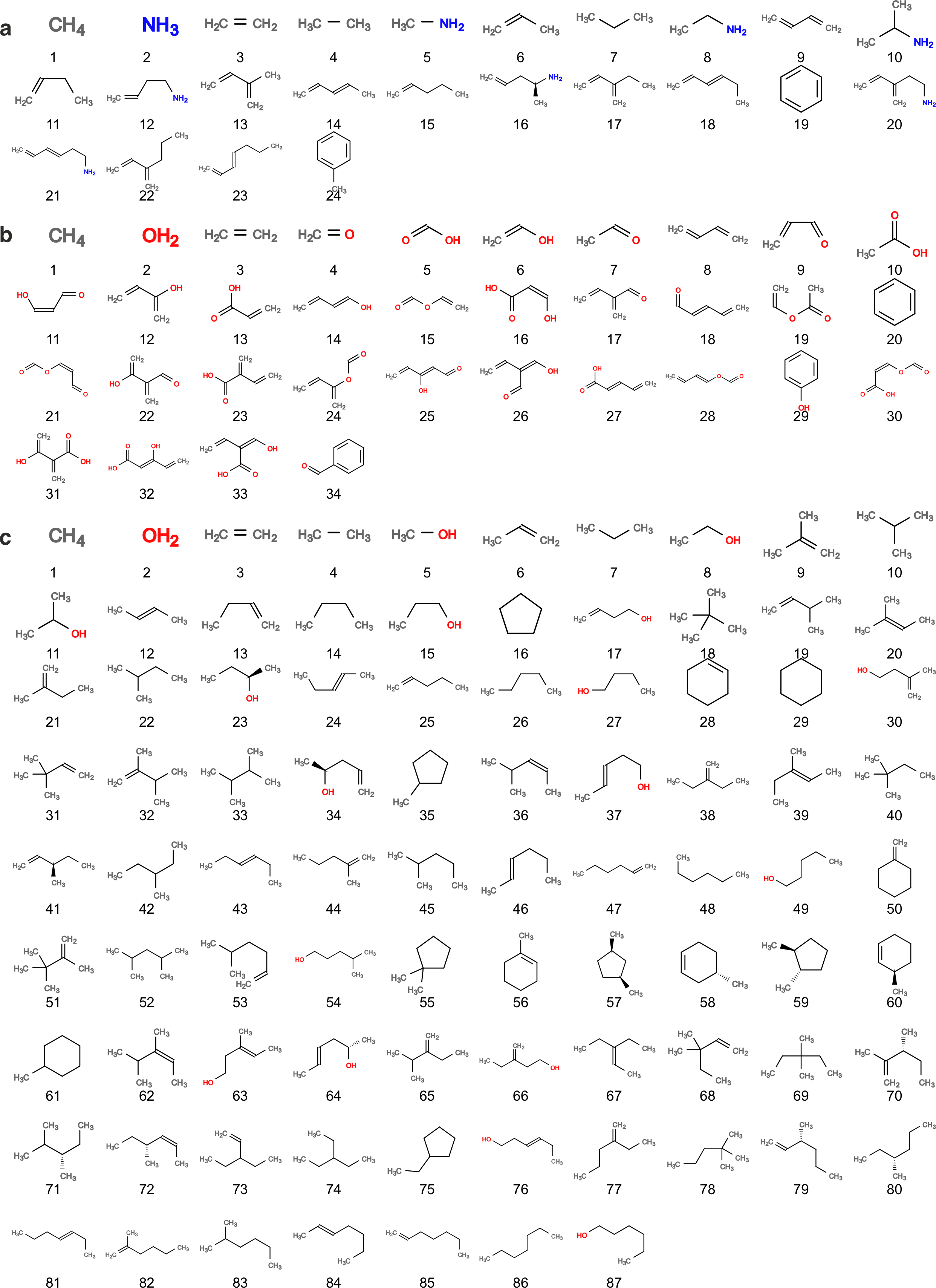}
		\caption{\label{sfig:bio_amons} 
            Amons of bio-moelcules, illustrated for amphetamine (A), aspirin (B) and cholesterol (C). For better visualization/understanding, only amon graphs are shown.}
	\end{figure}
	\clearpage

	\begin{figure}
		\centering
		\includegraphics[scale=0.74]{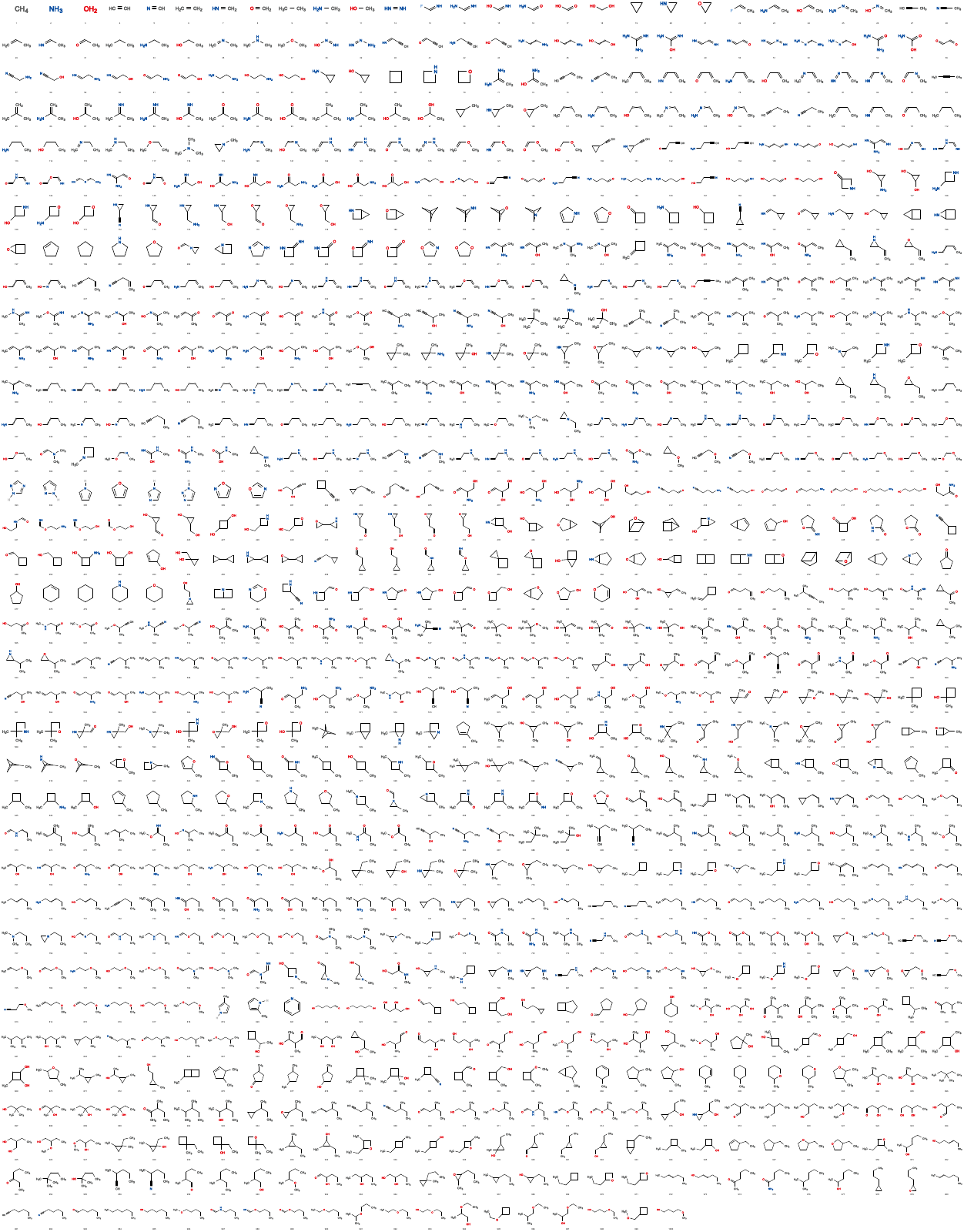}
		\caption{\label{sfig:qm9amons1k} 
			The 1k most frequent amons of QM9 molecules.}
	\end{figure}
	\clearpage
	
	\begin{figure}
		\centering
		\includegraphics[scale=0.5]{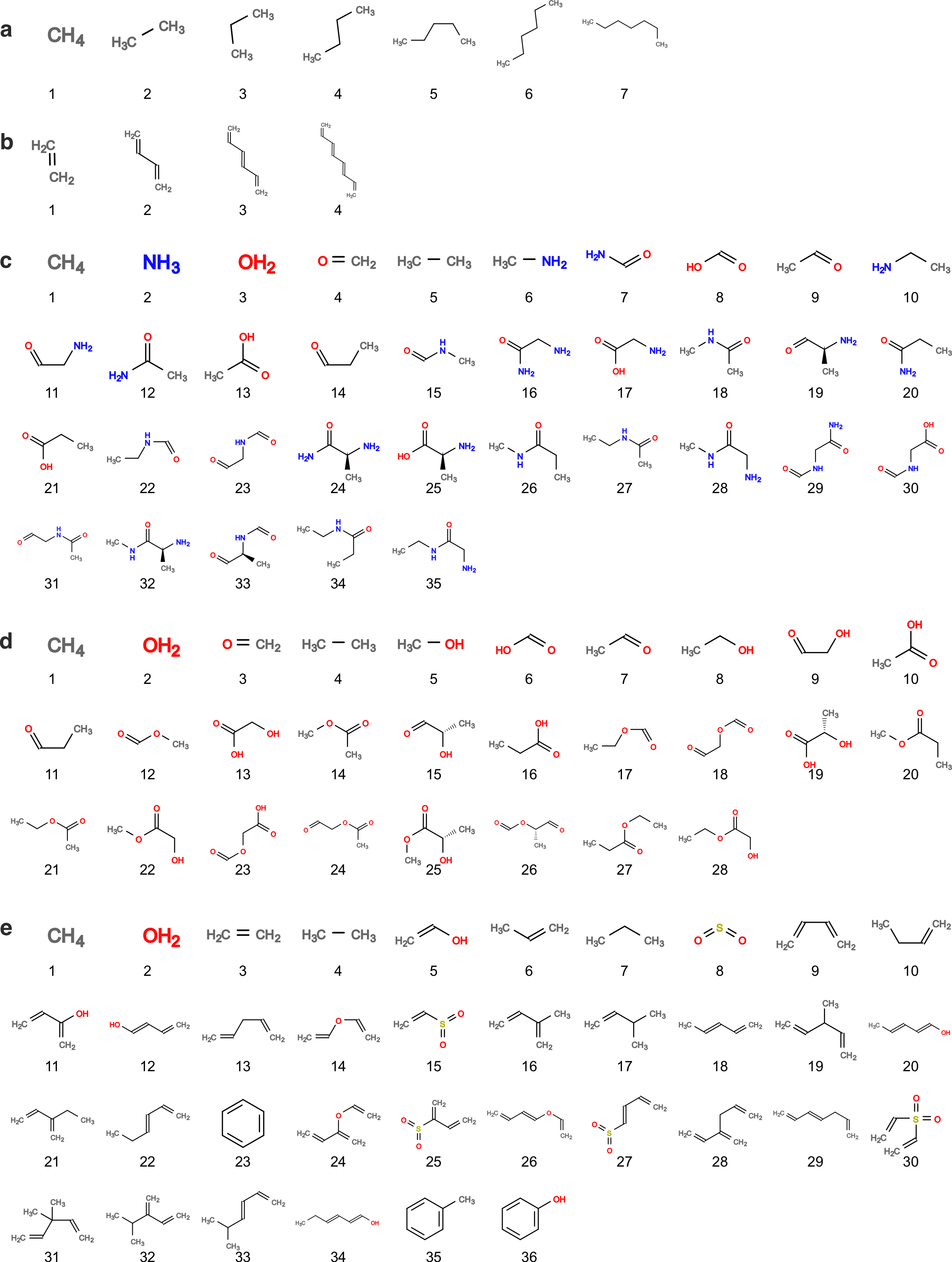}
		\caption{\label{sfig:polymers_amons} 
			Amons of polymers, illustrated for A: polyethylene (PE), B: polyacetylene (PA), C: alanine peptide ((ala)$_{10}$), D: polylactic acid (PLA) and E: the backbone of quaternary ammonium polysulphone (bQAPS). Anly amon graphs are shown for better visual effect.}
	\end{figure}
	\clearpage

	\begin{figure}
		\centering
		\includegraphics[scale=0.5]{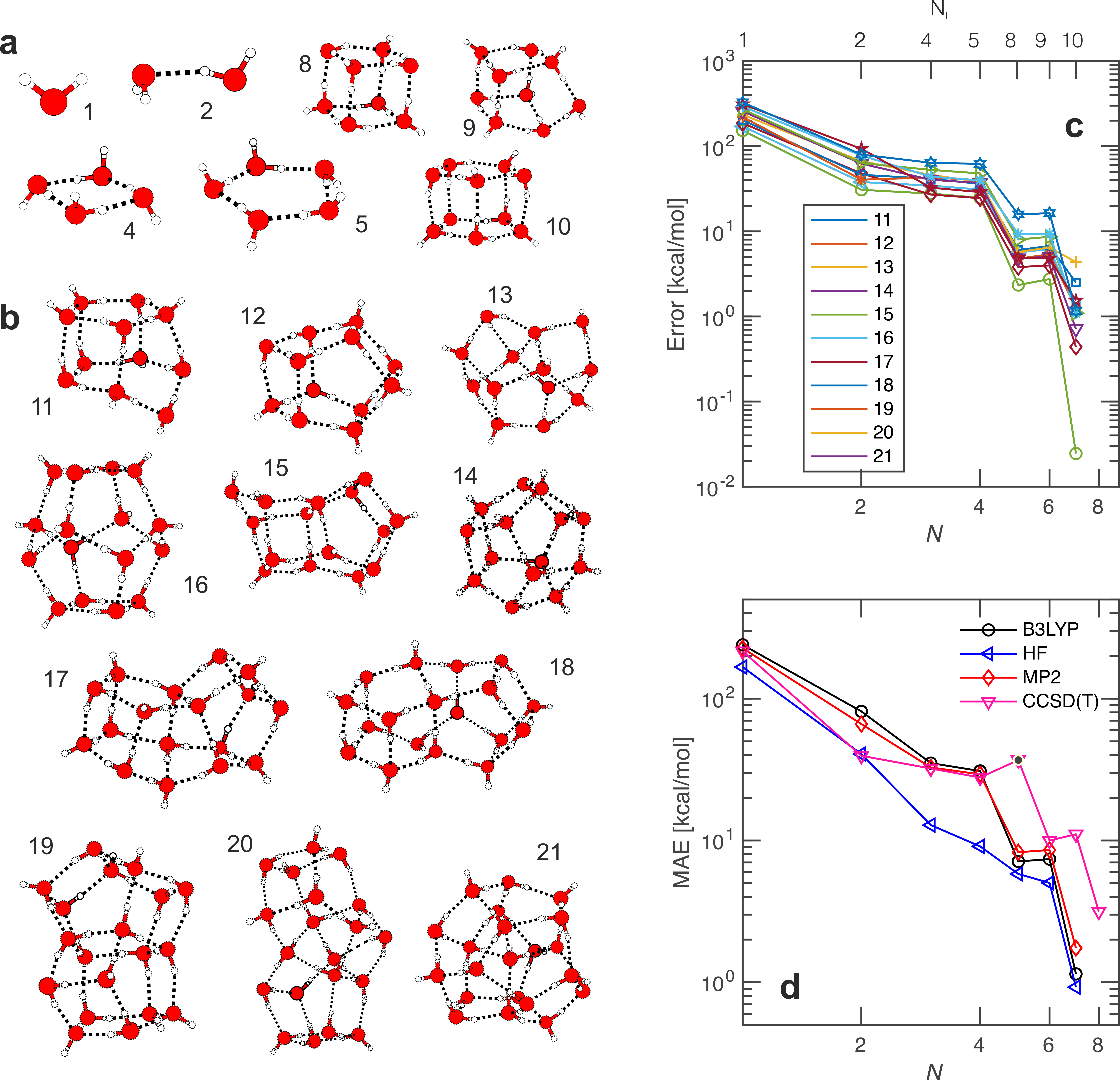}
		\caption{\label{sfig:water} 
			AML predictions for water cluster. (A) Small water clusters with $N_I\leq 10$, used as amons. Inset integer indicate the number of water molecules for each cluster; (B) query water clusters, ranging from (H$_2$O)$_{11}$ to (H$_2$O)$_{21}$; (C) Absolute error of predicted energies by AML as a function of amon number/size for each of the query water clusters in (B). Geometries and energies were obtained at the level of Gaussian 09/B3LYP/6-31G*; (D) Learning curve, reported as the mean absolute error (MAE) of total energy predictions by AML for all query water clusters as a function of number of amons. Reference data calculated at 4 levels of theory were used for training and test: HF (with HF coordinates, left-pointing black triangle), MP2 (with MP2 coordinates, red diamond), B3LYP (with B3LYP coordinates, black circle) and CCSD(T) (with MP2 coordinates, downward-pointing pink triangle) (all with the same Pople basis 6-31G*).  }
	\end{figure}
	\clearpage
	
	\begin{figure}
		\centering
		\includegraphics[scale=0.54]{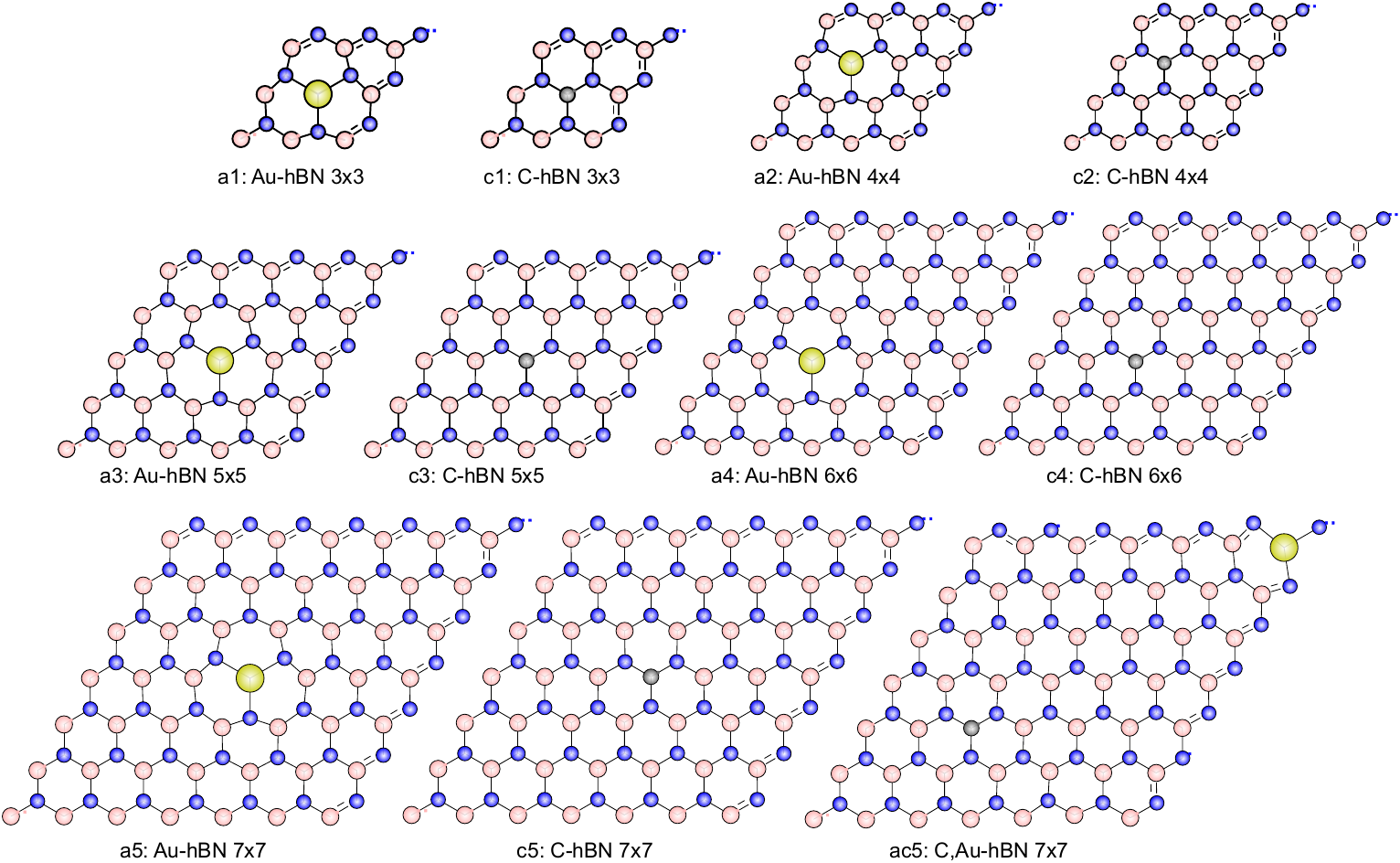}
		\caption{\label{sfig:hBN} 
			Amons for doped $h$BN sheets, illustrated for C-$h$BN 12x12 (corresponding amons include c1-c5), Au-$h$BN 12x12 (corresponding amons are a1-a5) and C,Au-$h$BN 12x12 (with all structures above being its amons). See Fig.~\ref{sfig:sca} for target structures.}
	\end{figure}
	\clearpage

	\begin{figure}
		\centering
        \includegraphics[scale=0.75]{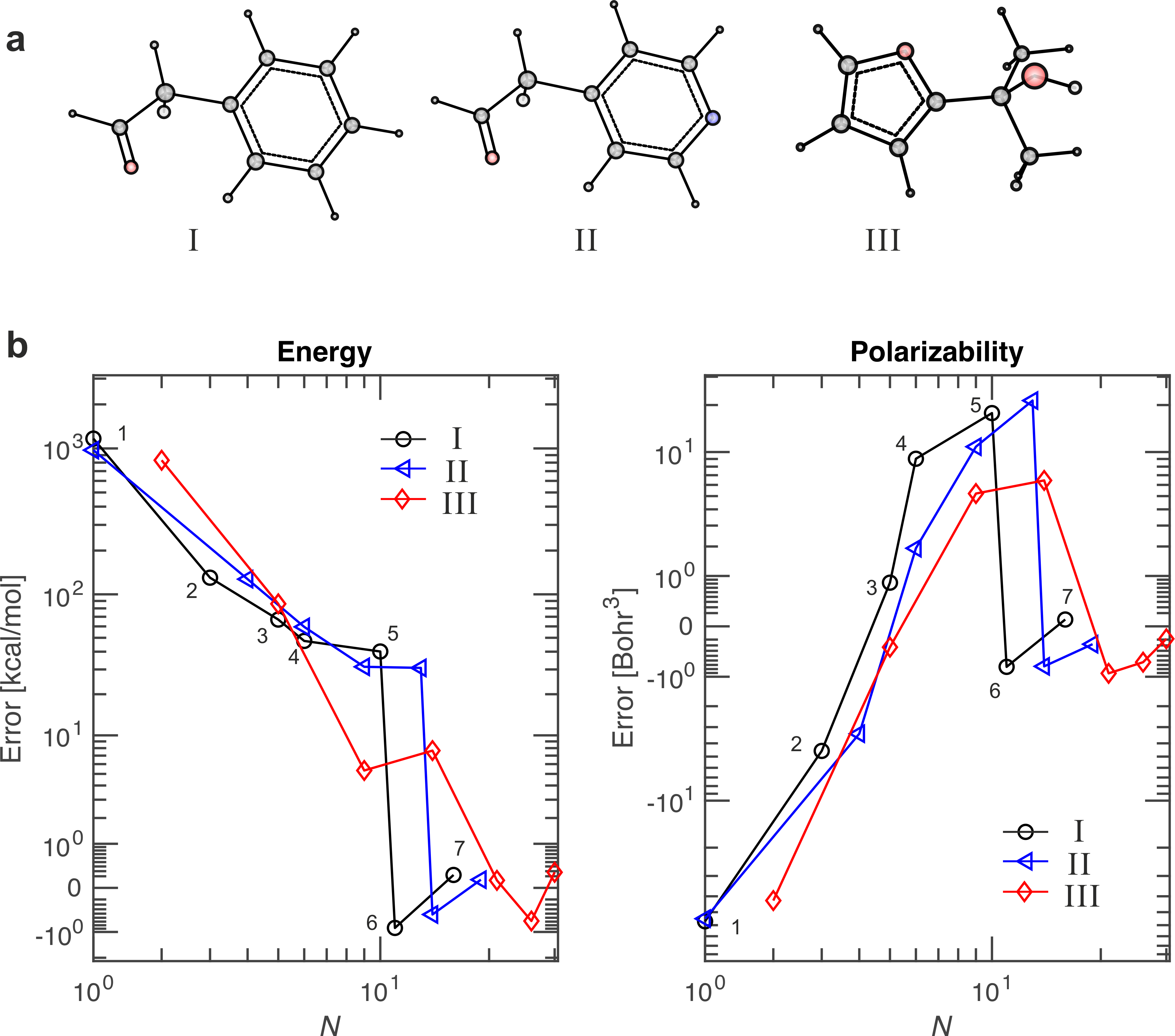}
		\caption{\label{sfig:EandAlphaFor3QM9} 
			AML predictions of energies and polarizabilities (B) for three exemplified query QM9 molecules with $N_I=9$ (A).
			Learning curves are shown as the difference beweetn AML-predicted values and corresponding DFT values, plotted  with respect to increasing number as well as size of amons. 
		}
	\end{figure}
	\clearpage
	
	\begin{figure}
		\centering
        \includegraphics[scale=0.7]{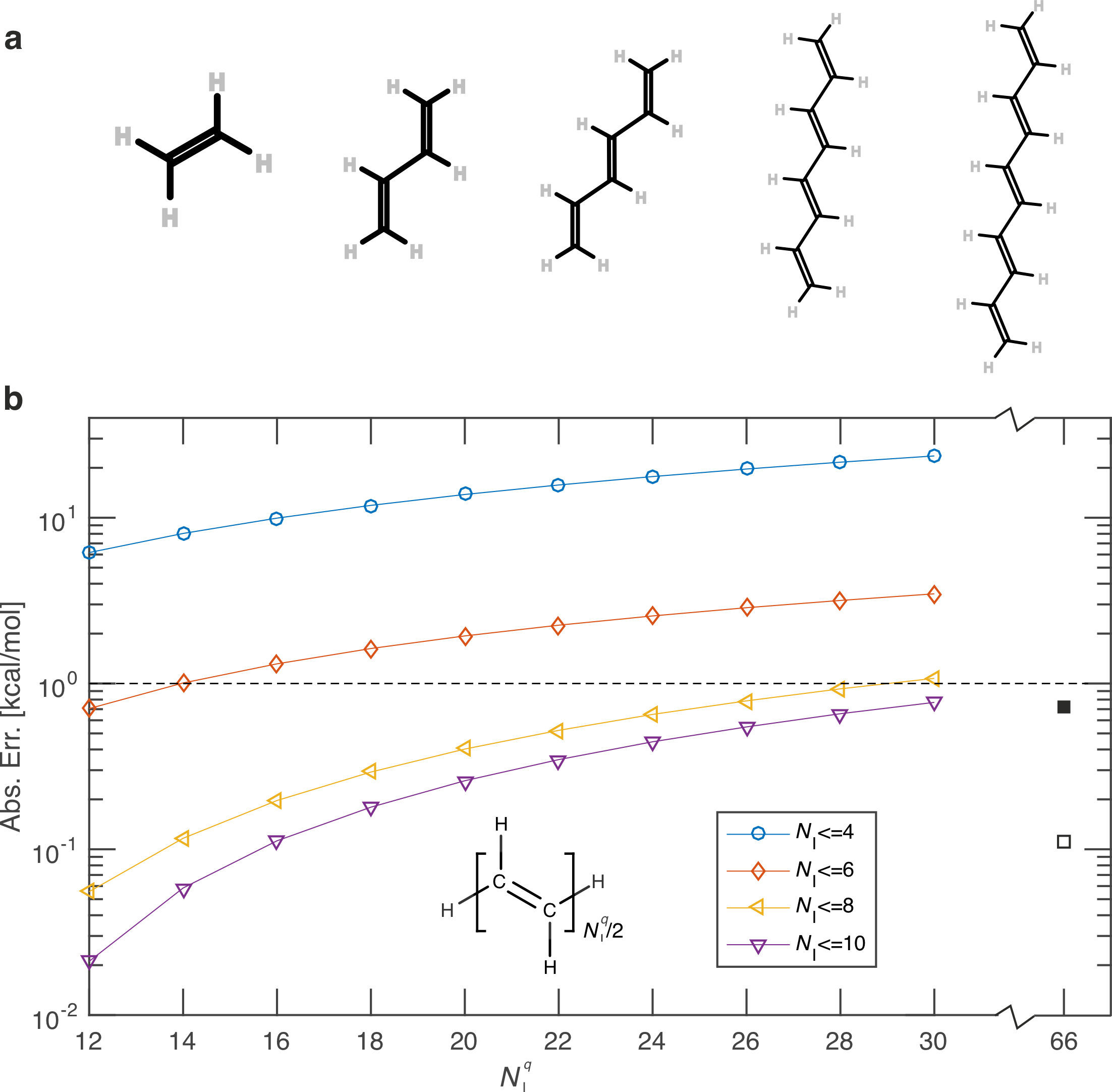}
		\caption{\label{sfig:conj} 
			Scalability of AML model applied to (trans) poly-acetylene (PA).
                        Panel (A) displays all amons employed with up to $N_I$ = 10 heavy atoms.
  (B) AML prediction error of atomization energy as a function of size of the query (for PA strands made from up to 15 acetylene monomers). 
 			Coloured symbols correspond to AML models trained on amons including increasingly larger instances 
			 with up to 4, 6, 8, or 10 heavy atoms, respectively. 
                        Squares show, for comparison, the error of a linear-scaling \emph{ab initio} method for PA with $N_I=56$ taken from Ref.~\cite{MTA_PA_2010} 
		}
	\end{figure}
	\clearpage

	\begin{figure*} 
		\centering 
		\includegraphics[scale=0.7]{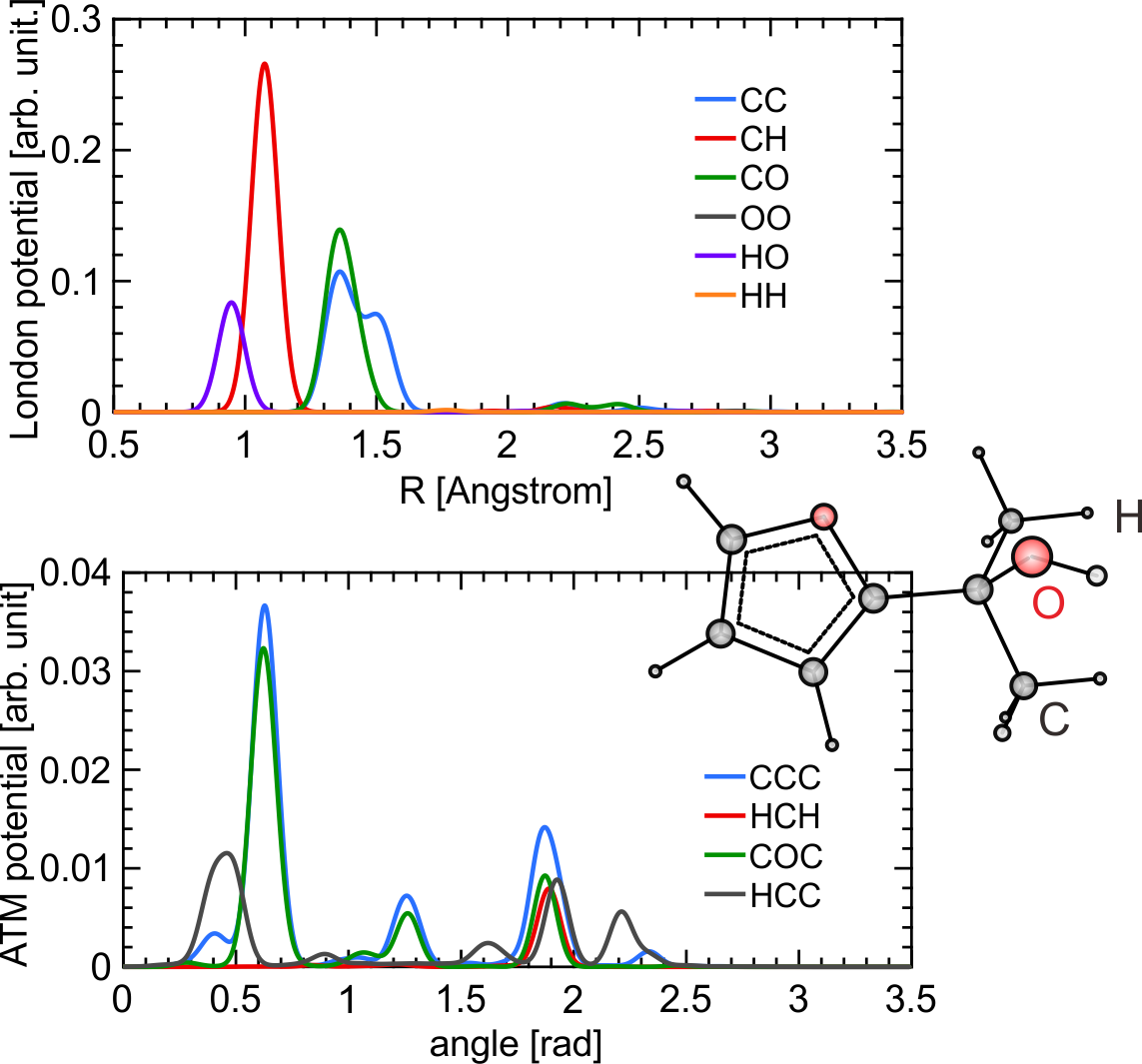}
		\caption{\label{sfig:SLATM1} 
			Spectrum of London potential (upper panel) and ATM potential (lower pannel) for the query molecule 2-(furan-2-yl)propan-2-ol shown in Fig.~\ref{fig:gdb9}A in the main text.}
	\end{figure*}
	\clearpage
	
	\begin{figure} 
		\centering
		\includegraphics[scale=0.5]{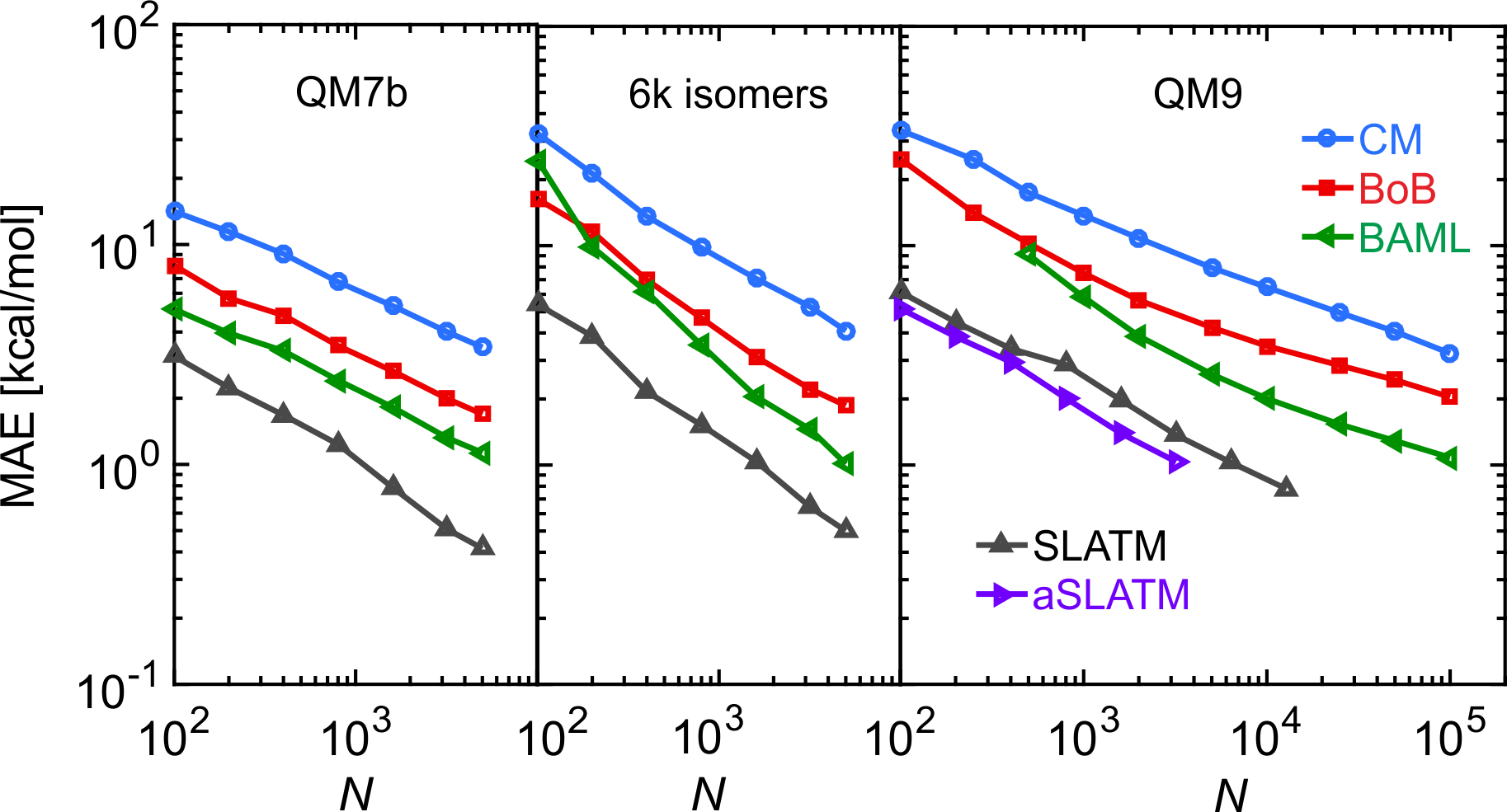}
		\caption{\label{sfig:SLATM2} 
			Comparison of learning curves obtained from four different representation based ML models (Coulomb matrix (CM)~\cite{CM}, Bag of Bond (BoB)~\cite{BoB}, Bond, Angle based ML (BAML)~\cite{baml}, and SLATM for the prediction of total molecular potential energy for three datasets: QM7b (left panel), QM9 (right panel) and 6k constitutional isomers from QM9 (middle panel).}
	\end{figure}
	\clearpage

\end{document}